\documentclass[acmsmall,screen,authorversion,nonacm]{acmart}

\usepackage{amsmath,amsfonts,bm}

\def\eqref#1{equation~\ref{#1}}

\def\1{\bm{1}}

\def\rvh{{\mathbf{h}}}

\def\rvs{{\mathbf{s}}}

\def\rvx{{\mathbf{x}}}
\def\rvy{{\mathbf{y}}}
\def\rvz{{\mathbf{z}}}

\def\rmI{{\mathbf{I}}}

\def\rmX{{\mathbf{X}}}

\def\vx{{\bm{x}}}

\def\evmu{{\mu}}

\def\evsigma{{\sigma}}

\DeclareMathAlphabet{\mathsfit}{\encodingdefault}{\sfdefault}{m}{sl}
\SetMathAlphabet{\mathsfit}{bold}{\encodingdefault}{\sfdefault}{bx}{n}

\usepackage{tablefootnote} 
\usepackage[flushleft]{threeparttable}

\AtBeginDocument{%
  \providecommand\BibTeX{{%
    \normalfont B\kern-0.5em{\scshape i\kern-0.25em b}\kern-0.8em\TeX}}}

\setcopyright{acmcopyright}
\copyrightyear{2022}
\acmYear{2022}
\acmDOI{XXXXXXX.XXXXXXX}

\acmJournal{JACM}
\acmMonth{9}

\usepackage{color} 
\usepackage{makecell}

\begin{document}

\title{3D Brain and Heart Volume Generative Models: A Survey}


\author{Yanbin Liu}
\email{csyanbin@gmail.com}
\orcid{0000-0003-4724-8065}
\affiliation{%
  \institution{Harry Perkins Institute of Medical Research, Department of Computer Science and Software Engineering, The University of Western Australia}
  \city{Canberra}
  \state{ACT}
  \country{Australia}
  \postcode{2601}
}

\author{Girish Dwivedi}\authornote{This work was supported by MRFF Frontier Health and Medical Research - RFRHPI000147.}
\affiliation{%
  \institution{Harry Perkins Institute of Medical Research, The University of Western Australia, Fiona Stanley Hospital}
  \city{Perth}
  \state{WA}
  \country{Australia}}
\email{girish.dwivedi@perkins.uwa.edu.au}

\author{Farid Boussaid}
\affiliation{%
  \institution{Department of Electrical, Electronic and
Computer Engineering, The University of Western Australia}
  \city{Perth}
  \state{WA}
  \country{Australia}}
\email{farid.boussaid@uwa.edu.au}

\author{Mohammed Bennamoun}
\affiliation{%
  \institution{Department of Computer Science and Software Engineering, The University of Western Australia}
  \city{Perth}
  \state{WA}
  \country{Australia}
}
\email{mohammed.bennamoun@uwa.edu.au}

\renewcommand{\shortauthors}{Yanbin L., Girish D., and Mohammed B.}

\begin{abstract}
Generative models such as generative adversarial networks and autoencoders have gained a great deal of attention in the medical field due to their excellent data generation capability.  
This paper provides a comprehensive survey of generative models for three-dimensional (3D) volumes, focusing on the brain and heart. 
A new and elaborate taxonomy of unconditional and conditional generative models is proposed to cover diverse medical tasks for the brain and heart: unconditional synthesis, classification, conditional synthesis, segmentation, denoising, detection, and registration. 
We provide relevant background, examine each task and also suggest potential future directions. 
A list of the latest publications will be updated on GitHub to keep up with the rapid influx of papers at \url{https://github.com/csyanbin/3D-Medical-Generative-Survey}.

\end{abstract}

\ccsdesc[500]{Computing methodologies~Computer vision}
\ccsdesc[500]{Computing methodologies~3D imaging}


\keywords{generative models, three-dimensional, medical images, brain and heart}

\maketitle

\section{Introduction}
A wide range of research fields has embraced deep learning (DL) in recent years, including image processing~\cite{ref:alexnet,ref:resnet,ref:mask_rcnn,ref:fast_rcnn}, speech recognition~\cite{ref:speech1,ref:speech2,ref:speech3}, natural language processing~\cite{ref:nlp1,ref:nlp2,ref:nlp3,ref:transformer,ref:bert}, and robotics~\cite{ref:robotics1,ref:robotics2}. 
Thus, the medical imaging community has put in significant efforts to take advantage of deep learning advances, and medical imaging research has made significant progress with respect to a variety of applications including classification~\cite{ref:cardiac_classification,ref:SCGANs,ref:3DCapsNet,ref:3DPixelCNN,ref:CRT_response}, segmentation~\cite{ref:3d_segmentation_survey,ref:gans_segmentation,ref:fcn_med_segment,ref:gans_segmentation2,ref:gans_segmentation}, registration~\cite{ref:Zhu_registration,ref:Ramon_registration}, detection~\cite{ref:Pinaya_detect,ref:Uzunova_detect,ref:diffusion_detect_segment}, denoising~\cite{ref:Wolterink_denoise,ref:LA-GANs,ref:3D_c-GANs,ref:RED-WGAN,ref:SGSGAN}, and synthesis~\cite{ref:3d_stylegan,ref:split_shuffle,ref:Liao_crossmodality,ref:Mcmt-gan,ref:ResViT}, as well as with various imaging modalities, including Computed Tomography (CT)~\cite{ref:CT1,ref:CT2}, ultrasound~\cite{ref:ultrasound}, Magnetic Resonance Imaging (MRI)~\cite{ref:MRI1,ref:MRI2}, and Positron Emission Tomography (PET)~\cite{ref:PET}.

A large number of annotated training images, obtained with the aid of crowd-sourcing annotation platforms like Amazon Mechanical Turk~\cite{ref:AMT}, were required for deep learning to be successful in natural image processing. However, the complexity of collection procedures, the lack of experts, privacy concerns, and the mandatory requirement of consent from patients make the annotation process a major bottleneck in medical imaging. 
In order to mitigate this issue, deep generative models (e.g., generative adversarial networks (GANs)~\cite{ref:gans} and variational autoencoder (VAE)~\cite{ref:VAE}) have been introduced to medical imaging. In these generative models, the original data distribution is mimicked so that realistic images are generated~\cite{ref:3d_stylegan,ref:slice_vae,ref:HA_GAN} or cross-modality synthesis can be achieved~\cite{ref:Lin_crossmodality,ref:Liao_crossmodality,ref:dEa-SA-GAN}.


There have been numerous survey papers published on deep generative models for medical imaging due to the rapid progress of the field~\cite{ref:gans_survey1,ref:gans_survey2,ref:gans_survey3,ref:gans_survey_aug,ref:gans_survey_seg1,ref:gans_survey_seg2,ref:survey_reg,ref:survey_MRI}. 
These surveys cover different medical applications and provide an overall review of GANs on general medical image analysis~\cite{ref:gans_survey1,ref:gans_survey2,ref:gans_survey3}. 
Some focus on a specific application only, such as augmentation~\cite{ref:gans_survey_aug}, segmentation~\cite{ref:gans_survey_seg1,ref:gans_survey_seg2} and registration~\cite{ref:survey_reg}. 
Others concentrate on a specific image modality, such as MRI~\cite{ref:survey_MRI}.
Even though many surveys exist, we find that there is a lack of comprehensive surveys on three-dimensional (3D) medical volume, which is the original data format of many medical modalities, such as MRI, CT, and PET. 
Moreover, existing surveys mainly focus on GANs, neglecting other effective generative models such as Autoencoders (AEs)~\cite{ref:AE1} and Autoregressive models~\cite{ref:pixelRNN}. 
In Section 2.1, we provide a comprehensive comparison to existing survey papers in the field of medical image generation (as shown in Table~\ref{tab:related_survey}) and detail what sets our survey apart.

\begin{figure}[t]
  \centering
  \includegraphics[width=0.94\textwidth]{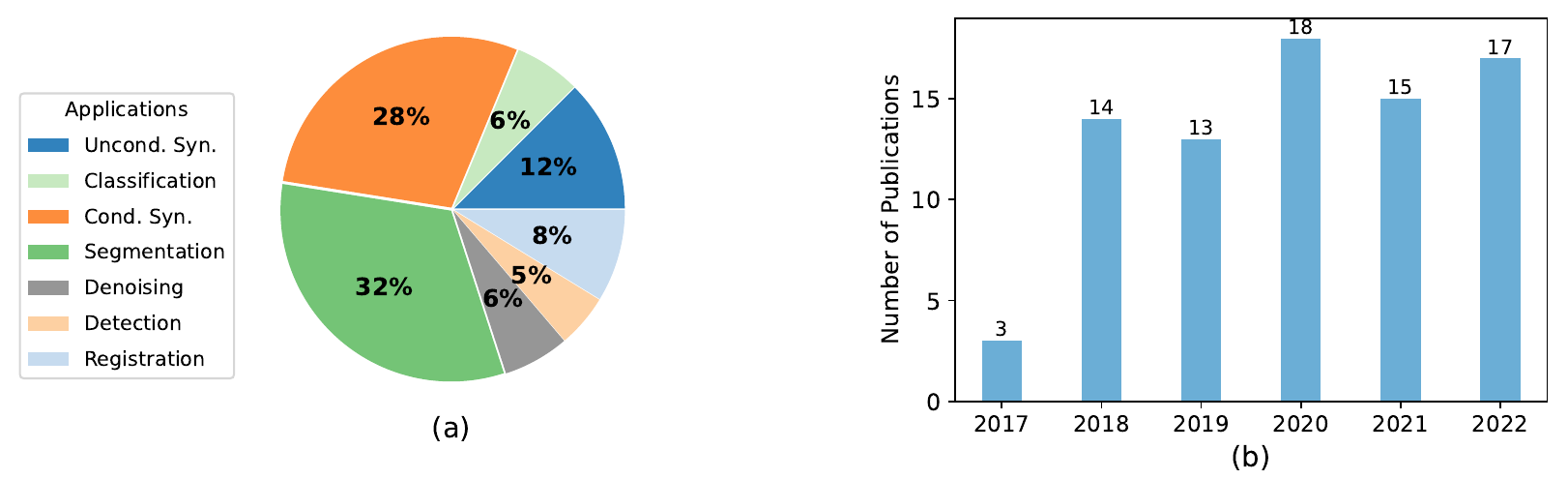}
  \caption{\label{fig:statistics}Statistics of the 3D brain and heart volume generative models. (a) Statistics of all publications according to medical applications. (b) Categorization by year of publication (2017-2022). Uncond. Syn.: Unconditional Synthesis, Cond. Syn.: Conditional Synthesis.}
\end{figure}

This inspired us to conduct a comprehensive survey of generative models for 3D medical volume images of the brain and heart. 
Since 3D volume is the intrinsic representation of many medical imaging modalities, it displays the entire and thorough anatomical structure of organs, whereas a 2D medical image only shows a specific view/plane. 
GANs are widely used for 3D medical volumes, but there has also been an increased interest in AEs (e.g., Diffusion Model~\cite{ref:diffusion_generation,ref:diffusion_detect_segment}) and Autoregressive models (e.g., Autoregressive Transformers~\cite{ref:Pinaya_detect}). 
As a result, our survey covers all three types of generative models. 
As far as organs are concerned, we restrict our interest to the brain and heart for the following reasons: \textbf{(1)} they are two vital organs that control the mental and physiological functions of the human body; \textbf{(2)} both organs involve a wide range of applications, e.g., segmentation (Fig.~\ref{fig:statistics}(a)); \textbf{(3)} generative models are essential for both organs because of their data scarcity; \textbf{(4)} by covering these 2 organs, we are able to cover generative models for both static (i.e., brain) and dynamic organs (i.e., heart).

\begin{figure}[t]
  \centering
  \includegraphics[width=0.99\textwidth]{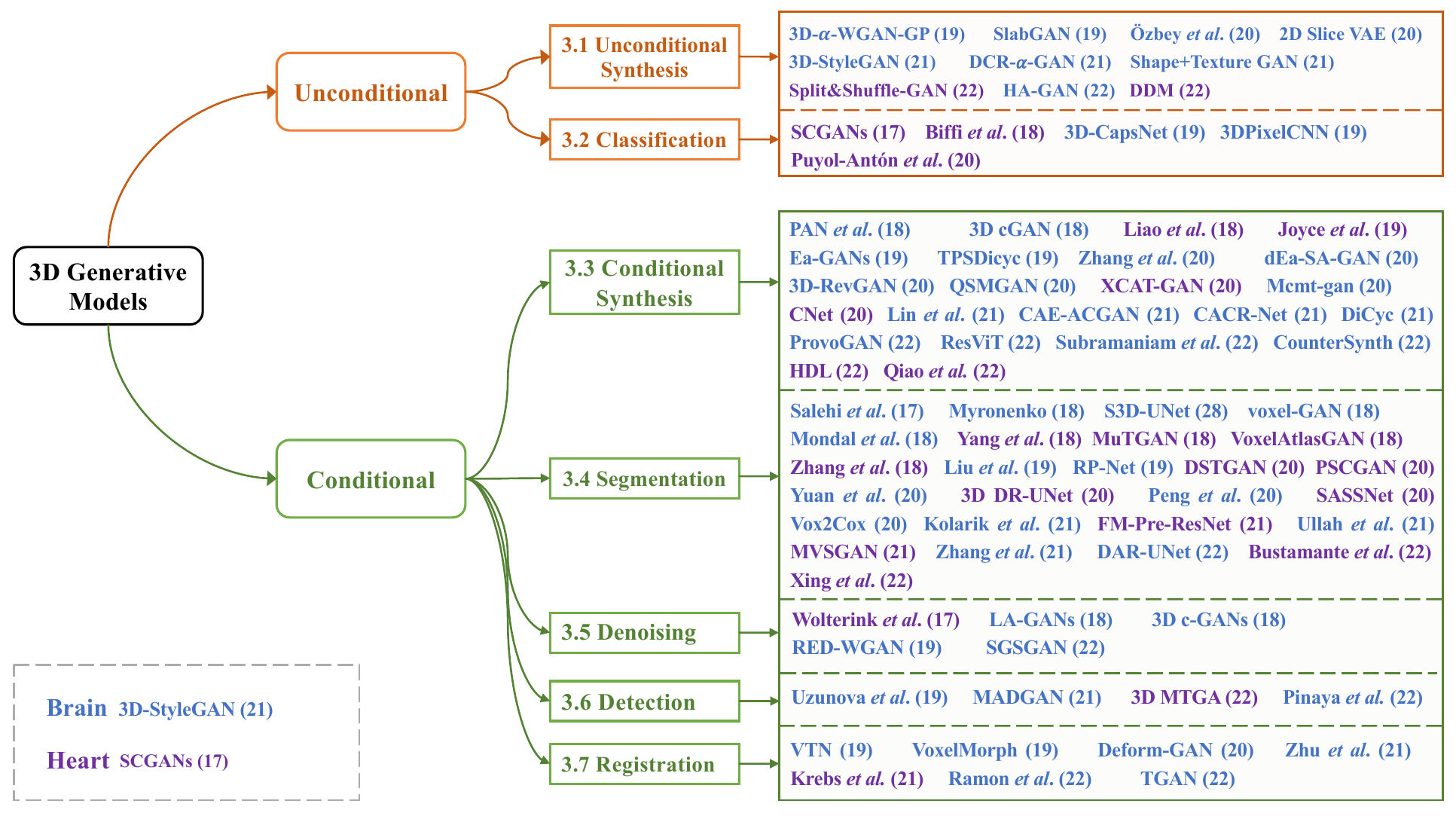}
  \caption{\label{fig:taxonomy}Proposed taxonomy of 3D generative models for the brain and the heart. The numbers in the parentheses denote the publication year (after 2000).}
\end{figure}

\textbf{Contributions.} 
To provide a comprehensive and organized survey, we introduce a new taxonomy (Fig.~\ref{fig:taxonomy}) that divides generative models into unconditional (only taking a random variable as input) and conditional (taking an additional data modality as input). 
In Figure~\ref{fig:statistics}, we provide a statistical analysis of the proportion of publications per application and the number of publications per year. Our contributions can be summarized as follows:
\begin{itemize}
    \item This is the \emph{first survey} on generative models for 3D medical volume images, focusing on two important organs, i.e., the brain and the heart. It aims to bridge the gap between the research of the 3D generative models community and the research of the medical imaging community.
    \item We provide a new taxonomy (Fig.~\ref{fig:taxonomy}) of 3D generative models by categorizing them as unconditional or conditional generative models. Every category includes several relevant medical applications. 
    \item Whilst most existing surveys focus on GANs only, we cover three main categories of generative models: GANs, AEs, and Autoregressive models. 
    \item We discuss the key challenges and future directions of 3D medical generative models and applications. 
\end{itemize}

\textbf{Paper Organization.} 
The remainder of this paper is organized as follows. 
We introduce the foundational techniques and challenges of 3D generative models in Section 2. 
Section 3 comprehensively elaborates on the 3D medical applications of both the unconditional and conditional generative models, including unconditional synthesis, classification, conditional synthesis, segmentation, denoising, detection, and registration. 
Then, Section 4 discusses the above-surveyed applications and gives four future directions. 
Finally, Section 5 concludes the paper.

\section{Background}

\subsection{Related Survey}


\begin{table}[ht]
\centering
\setlength{\tabcolsep}{3pt}
\caption{\label{tab:related_survey}Comparison with existing survey papers on medical image generation.}
\resizebox{1.0\textwidth}{!}{
\begin{tabular}{l|l|l|l|l|l} \hline
Publication & Year  &  Model(s) & Organ(s) & Image Format & Application(s)  \\ \hline 
Yi \emph{et al}.~\cite{ref:gans_survey2} & 2019 & GANs & All & mainly 2D, 3D & synthesis, reconstruction, segmentation, classification \\ 
&  &  &  &  & detection, registration \\ \hline
Kazeminia \emph{et al}.~\cite{ref:gans_survey3} & 2020 & GANs & All & mainly 2D, 3D & synthesis, segmentation, reconstruction, detection \\ 
&  &  &  &  & de-noising, registration, classification \\ \hline
ALAMIR \emph{et al.}~\cite{ref:gans_survey1} & 2022 & GANs & All & mainly 2D, 3D & cross-modality, segmentation, augmentation, reconstruction\\
& & & & & detection, classification, registration \\ \hline
Chen \emph{et al.}~\cite{ref:gans_survey_aug} & 2022 & GANs & All & 2D, 3D & augmentation \\\hline
Iqbal \emph{et al.}~\cite{ref:gans_survey_seg1} & 2022 & GANs & All & 2D, 3D & segmentation \\\hline
Jeong \emph{et al.}~\cite{ref:gans_survey_seg2} & 2022 & GANs & All & mainly 2D, 3D & classification, segmentation \\\hline
Ali \emph{et al.}~\cite{ref:survey_MRI} & 2022 & GANs & Brain & 2D, 3D & only statistics of all applications (no description) \\\hline
This Survey & 2023 & GANs, AEs & Brain, Heart & focus on 3D & unconditional synthesis, classification, conditional synthesis \\ 
&  & Autoregressive &  &  & segmentation, denoising, detection, registration \\ \hline
\end{tabular}  
}
\end{table}

As generative models find increasing use in the medical field, many survey papers have been published to provide overviews~\cite{ref:gans_survey2,ref:gans_survey3,ref:gans_survey1,ref:gans_survey_aug,ref:gans_survey_seg1,ref:survey_MRI}. 
In Table~\ref{tab:related_survey}, we differentiate our survey from existing ones by comparing key elements such as Model(s), Organ(s), Image Format, and Application(s). The distinct contributions and unique aspects of our survey are summarized below: 
\begin{itemize}
\item 
While the majority of existing surveys concentrate on Generative Adversarial Networks (GANs), our survey includes all three major types of generative models: GANs, Autoencoders (AEs), and Autoregressive models. Notably, a recent AE variant, Denoising Diffusion Probabilistic Models (DDPMs)~\cite{ref:DDPM}, has outperformed GANs in the generation of natural images~\cite{ref:beat_gans,ref:beat_gans2}. Additionally, Autoregressive Transformer~\cite{ref:autoregressive_transformer} has shown promise in generating high-resolution images. Both AEs and Autoregressive models are poised to make significant contributions to medical image generation in the near future.

\item 
We observed that there is a gap in the literature when it comes to survey papers focused specifically on brain and heart image generation. While Ali \emph{et al.}~\cite{ref:survey_MRI} do cover brain MRI, their scope is limited to providing general statistics on demographics, applications, evaluations, and datasets. To our knowledge, no survey exists that focuses on heart image generation. Our survey aims to fill this gap.

\item 
Given that GANs were initially developed for 2D images, several existing surveys~\cite{ref:gans_survey2,ref:gans_survey3,ref:gans_survey1,ref:gans_survey_seg2} primarily focus on 2D generative models. In contrast, our survey focuses on 3D volumes, the native image format for medical imaging. Utilizing this 3D format offers additional advantages for subsequent applications~\cite{ref:3d_stylegan,ref:split_shuffle}.

\end{itemize}

\subsection{Unconditional and Conditional Generative Models}

\begin{figure}[t]
  \centering
  \includegraphics[width=0.8\textwidth]{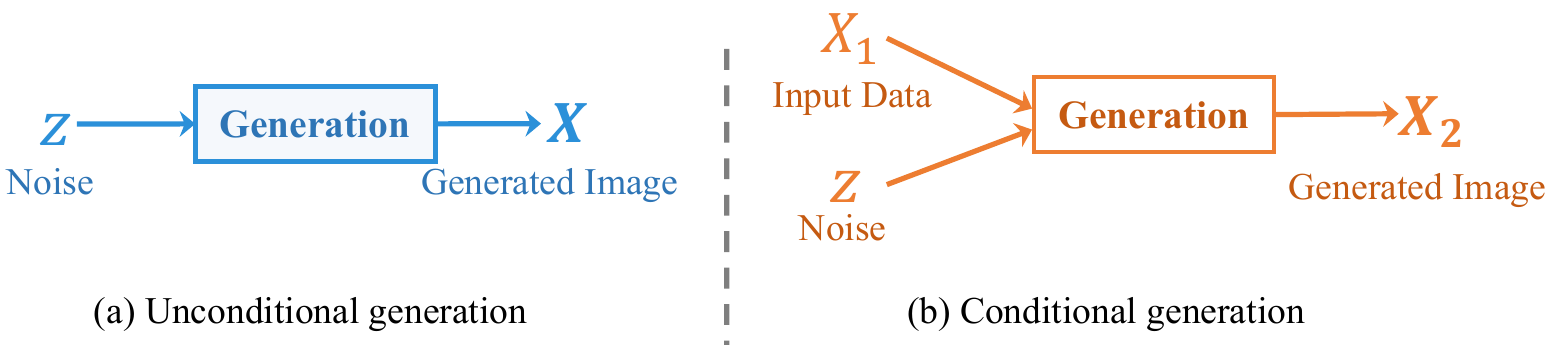}
  \caption{\label{fig:condition}Generation process in unconditional and conditional generative models.
  }
\end{figure}

In this paper, we divide generative models for 3D volume images into two categories: unconditional model and conditional model (Fig.~\ref{fig:condition}), where we only show the generation process for simplicity. 

In an unconditional generative model (Fig.~\ref{fig:condition}(a)), the input is a random noise variable $z$, and the output from the generation model is the generated image $\textbf{X}$. Several model architectures belong to this type of model. 
A GAN generator, for example, only uses random noise variables to synthesize images. 
Random Gaussian variables are input to the decoder of the VAE model. 
The unconditional generative model is used in several 3D medical applications, including unconditional synthesis~\cite{ref:3D_a_WGANGP,ref:3d_stylegan,ref:split_shuffle,ref:slice_vae} and classification~\cite{ref:SCGAN,ref:3DPixelCNN,ref:cardiac_classification}. 

In a conditional generative model (Fig.~\ref{fig:condition}(b)), in addition to the random noise $z$, the informative input data $X_1$ (e.g., semantic or visual input) is also fed to the generation model to help generate the output image $\rmX_2$. 
Depending on the applications, a variety of data formats are supported by $X_1$, including class labels~\cite{ref:cond_gans}, attributes~\cite{ref:cond_gans_attr}, texts~\cite{ref:cond_gans_text}, and images~\cite{ref:image_to_image}. 
For example, the original cGAN~\cite{ref:cond_gans} generated synthetic MNIST~\cite{ref:mnist} images conditioned on the class labels. 
Pixel-to-pixel~\cite{ref:image_to_image} achieved image style transfer by training a conditional GAN whose generator and discriminator were both based on the input images. 
Many 3D medical applications take advantage of the conditional generative model, including conditional synthesis~\cite{ref:CounterSynth,ref:3DcGAN,ref:PAN,ref:Liao_crossmodality,ref:Ea-GANs,ref:CAE-ACGAN,ref:ResViT}, segmentation~\cite{ref:voxel-GAN,ref:Vox2Vox,ref:Zhang_segmentation,ref:voxel-GAN,ref:Zhang_segment2}, denoising~\cite{ref:Wolterink_denoise,ref:LA-GANs,ref:3D_c-GANs,ref:RED-WGAN,ref:SGSGAN}, detection~\cite{ref:Uzunova_detect,ref:MADGAN,ref:3D_MTGA,ref:Pinaya_detect}, and registration~\cite{ref:VTN,ref:VoxelMorph,ref:Deform-GAN,ref:Ramon_registration,ref:TGAN}. 

From the data distribution perspective, the unconditional generative model captures the original realistic data distribution without requiring any additional information other than random noise. 
The conditional generative model can be regarded as a transformation from the input data distribution $p(\rmX_1)$ to the output image distribution $p(\rmX_2)$. 
By proposing a new unconditional/conditional taxonomy perspective, we hope researchers can gain valuable insights into future model design and apply them to target medical applications in the future.

\subsection{Generative Adversarial Networks}

Generative Adversarial Networks were proposed by Goodfellow~\emph{et al.}~\cite{ref:gans} in 2014. The main idea is to design two networks (a discriminator and a generator) to contest with each other in a zero-sum two-layer game. Specifically, the generator $G$ takes a random noise variable $z$ as the input to generate the synthesized images $G(z)$. 
The role of the discriminator is to distinguish between the realistic images $x$ and the generated fake images $G(z)$. In an ideal case, the two-player game reaches a Nash equilibrium~\cite{ref:nash1,ref:FID} where the synthesized images $G(z)$ are indistinguishable from real images $x$. 
The equilibrium is difficult to achieve in practice, and GANs training suffers from two problems, i.e., training instability~\cite{ref:gans_instability1,ref:gans_instability2} and mode collapse~\cite{ref:gans_mode1,ref:gans_mode2}. 
Diverse architectures and training strategies have been proposed to address the two problems and improve GANs performance~\cite{ref:gans_improved1,ref:gans_improved2,ref:gans_improved3,ref:gans_improved4}. Below, we describe the variants that are most relevant to the generation of 3D medical volume images.  


\subsubsection{Vanilla GAN}
\begin{figure}[t]
  \centering
  \includegraphics[width=0.8\textwidth]{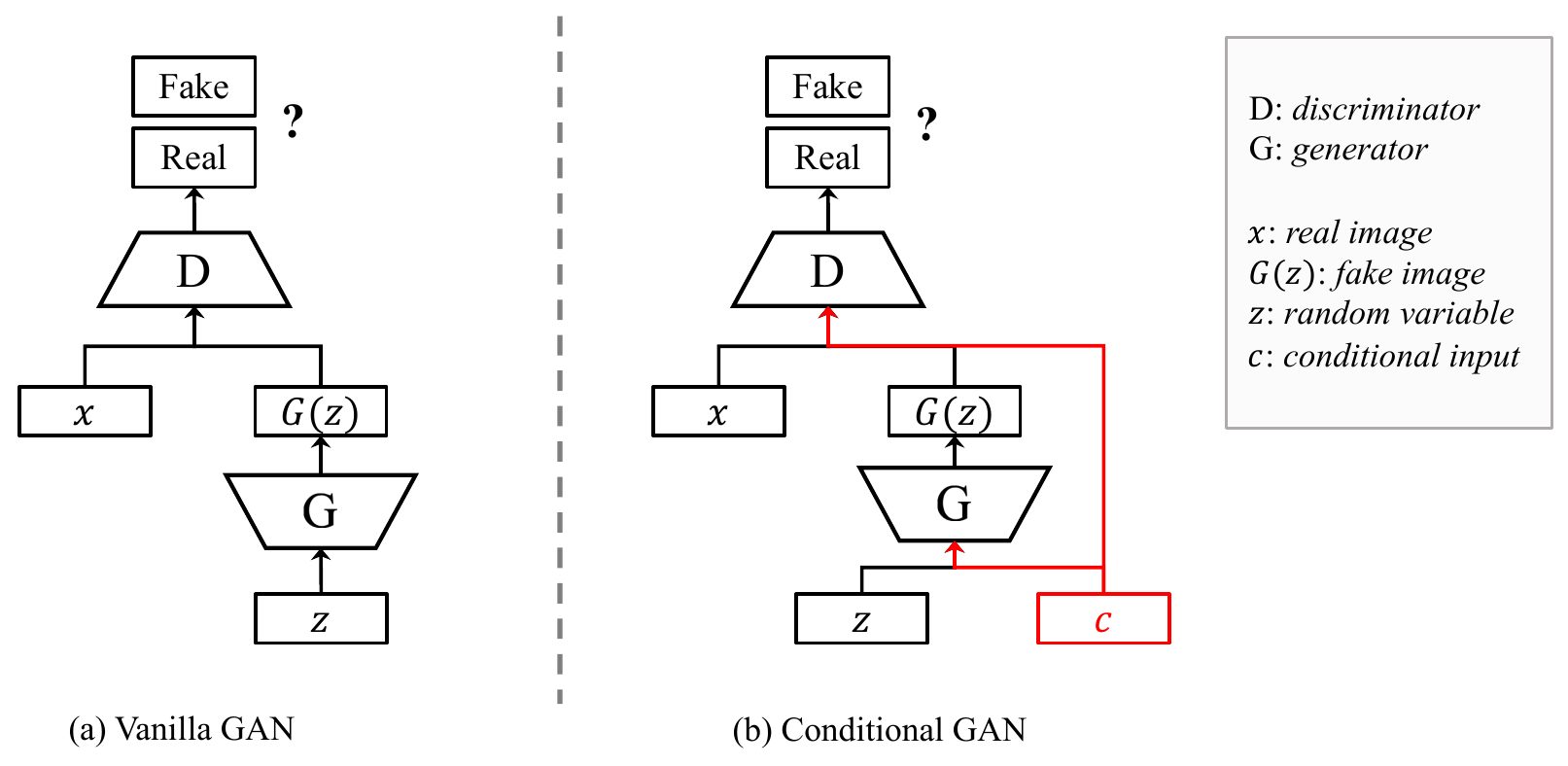}
  \caption{\label{fig:condGAN}The structure of (a) the vanilla (unconditional) and (b) the conditional GAN.}
\end{figure}

The structure of the vanilla GAN is shown in Fig.~\ref{fig:condGAN}(a). 
Based on a prior distribution of the input noise variable $z\sim p_z(z)$, the generator trains its distribution $p_g$ over $x$ to approximate the real data distribution $p_{data}$. The discriminator maximizes the accuracy of classifying real/fake images by optimizing over $D(x)$/$D(G(z))$. The minmax optimization problem for $G$ and $D$ is defined as follow:
\begin{equation}
    \label{eq:gan}
    \min_{G} \max_{D} V(D, G) = \mathbb{E}_{x\sim p_{data}(x)} [\log D(x)] + \mathbb{E}_{z\sim p_z(z)}[\log (1-D(G(z)))]\,. 
\end{equation}
Theoretically, for arbitrary functions $G$ and $D$, the optimal solution satisfies $p_g = p_{data}$ and $D(x) = D(G(z)) = \frac{1}{2}$. 
But in practice, $G$ and $D$ are usually implemented by deep neural networks (e.g., multilayer perceptrons or convolution neural networks), which only have limited capacity and cover a limited family of $p_g$ distributions. 

\subsubsection{Conditional GAN}
The vanilla GAN is unconditional, which means it cannot use additional information or control the modes of the generated data. 
Hence, conditional GAN (cGAN)~\cite{ref:cond_gans} was proposed in order to improve the flexibility of vanilla GAN by conditioning on additional information. As shown in Fig.~\ref{fig:condGAN}(b), the conditional input $c$ is fed into both the generator $G$ and discriminator $D$ (red line). The minmax optimization problem for $D$ and $G$ is then defined as follow:
\begin{equation}
    \label{eq:cond_gan}
    \min_{G} \max_{D} V(D, G) = \mathbb{E}_{x\sim p_{data}(x)} [\log D(x|c)] + \mathbb{E}_{z\sim p_z(z)}[\log (1-D(G(z|c)))]\,. 
\end{equation}
The conditional GAN (Eq.~\ref{eq:cond_gan}) is different from the unconditional GAN (Eq.~\ref{eq:gan}) because it conditions on $c$ in the generation and discrimination process. 
$c$ is used in the original paper~\cite{ref:cond_gans} to represent labels in the MNIST~\cite{ref:mnist} image generation and to represent image features in the generation of multi-model Flickr~
tags. 

The conditional input $c$ can be instantiated with different information and modalities so as to perform various image generation tasks. These tasks include text-to-image synthesis~\cite{ref:stackgan,ref:stackgan++}, image-to-image translation~\cite{ref:image_to_image,ref:un_image_to_image}, attribute editing~\cite{ref:cond_gans_attr,ref:cond_gans_attr2}, image segmentation~\cite{ref:gans_segmentation,ref:gans_segmentation2}, etc. 

%

\subsubsection{Pixel-to-pixel} 
Pixel-to-pixel was proposed in~\cite{ref:image_to_image} to solve the general image-to-image translation problems, e.g., edge to photo, day to night. 
In its implementation, both the generator and discriminator are fed with input images to construct a conditional GAN architecture. The objective function is as follows:
\begin{equation}
    \min_{G} \max_{D} \mathbb{E}_{x,y} [\log D(x,y)] +\mathbb{E}_{x,z} [\log (1 - D(x, G(x,z)))] + \lambda \mathbb{E}_{x,y,z} [\|y-G(x,z)\|_1]\,.
\end{equation}
Pixel-to-pixel is trained with image pairs $(x,y)$, and translates images from $x$ domain to $y$ domain after training. 


\subsubsection{CycleGAN} 
Since Pixel-to-pixel~\cite{ref:image_to_image} relies on a large number of annotated image pairs, it is time-consuming and sometimes infeasible. 
CycleGAN~\cite{ref:un_image_to_image} tackles this problem by taking advantage of the concept of cycle-consistency. Two discriminators $G$ and $F$ are designed for forward and backward mapping, and two discriminators $D_X$ and $D_Y$ are devised for adversarial learning. The objective function is as follows:
\begin{equation}
    L(G, F, D_X, D_Y) = L_{GAN}(G, D_Y, X, Y) + L_{GAN}(F, D_X, Y, X) + \lambda L_{cyc}(G, F)\,,
\end{equation}
where $L_{GAN}$ is the adversarial loss, and $L_{cyc}$ is the cycle-consistency loss.


\subsubsection{WGAN and WGAN-GP} 
The GANs training objective can be non-continuous with respect to the generator parameters, resulting in training instability. 
To alleviate this issue, Wasserstein GAN (WGAN)~\cite{ref:wgan} was proposed to leverage the Wasserstein distance as its objective function. 
To further improve the training stability and generation quality, \cite{ref:wgan_gp} designed WGAN-GP, a GAN variant, by adding the gradient penalty in the WGAN training loss. 
The WGAN-GP objective function is:
\begin{equation}
    L = \mathbb{E}_{\tilde{\rvx} \sim p_g} [D(\tilde{\rvx})]  -  \mathbb{E}_{\rvx \sim p_{data}} [D(\rvx)] +\lambda \mathbb{E}_{\hat{\rvx}\sim p_{\hat{\rvx}}} [(\|\nabla_{\hat{\rvx}} D(\hat{\rvx}) \|_2 - 1)^2]  \,,
\end{equation}
where $\hat{\rvx}$ are random samples used to calculate the gradient.

\subsubsection{StyleGAN and StyleGAN2} 
A style-based generative model, such as StyleGAN~\cite{ref:stylegan}, uses a mapping network to produce style vectors for each convolution layer of the generator. In StyleGAN, the Adaptive Instance Normalization (AdaIN) operation takes the style vectors as input to control the generation style. 
With this method, diverse and high-quality images can be generated with good interpolation properties. For example, style vectors corresponding to specific attributes can be jointly utilized to generate images of desired attributes, such as pose, skin tone, and hairstyle. 
However, StyleGAN suffers from characteristic artifacts (e.g., water droplet-like artifacts), which are corrected by the improved StyleGAN2~\cite{ref:stylegan2} model. 
By design, StyleGAN2 designed the modulation and demodulation modules to incorporate style vectors into the convolution operation. Moreover, it proposed lazy regularization and alternative network designs (i.e., input/output skips and residual nets). 
As a result of its good performance, the StyleGAN2 model has been used in a wide range of medical imaging methods~\cite{ref:split_shuffle,ref:3d_stylegan}.

\subsection{Autoencoders}
The concept of autoencoder dates back to the early stages of neural networks~\cite{ref:AE1,ref:AE2,ref:AE3}. 
The original purpose of AEs was to perform dimensionality reduction or feature learning in an unsupervised manner. 
An autoencoder is composed of an encoder $E$ with parameter $\phi$ and a decoder $D$ with parameter $\theta$. Typically, an autoencoder has a bottleneck structure (a.k.a undercomplete), i.e., the latent embedding dimension of the encoder is smaller than the original data $x$. 
With this design, the autoencoder can capture the intrinsic features of the data. 
The training of an autoencoder can be achieved by minimizing the reconstruction error (e.g., mean squared error) as follows:
\begin{equation}
    \label{eq:AE}
    \min_{\theta,\,\phi} \mathbb{E}_x\, \|x - D_\theta(E_\phi(x))\|^2 \,.
\end{equation}

With the recent emergence of deep neural networks, a multitude of AE model variants have been developed, such as Denoising Autoencoder (DAE)~\cite{ref:DAE,ref:DAE2}, VAE~\cite{ref:VAE}, conditional VAE~\cite{ref:CVAE,ref:CVAE2}, VQVAE~\cite{ref:VQVAE}, and U-Net~\cite{ref:UNET}. 
We cover below the most relevant AE models for the synthetic generation of 3D medical volume images.

\subsubsection{Variational Autoencoder (VAE)} 
\begin{figure}[ht]
  \centering
  \includegraphics[width=0.6\textwidth]{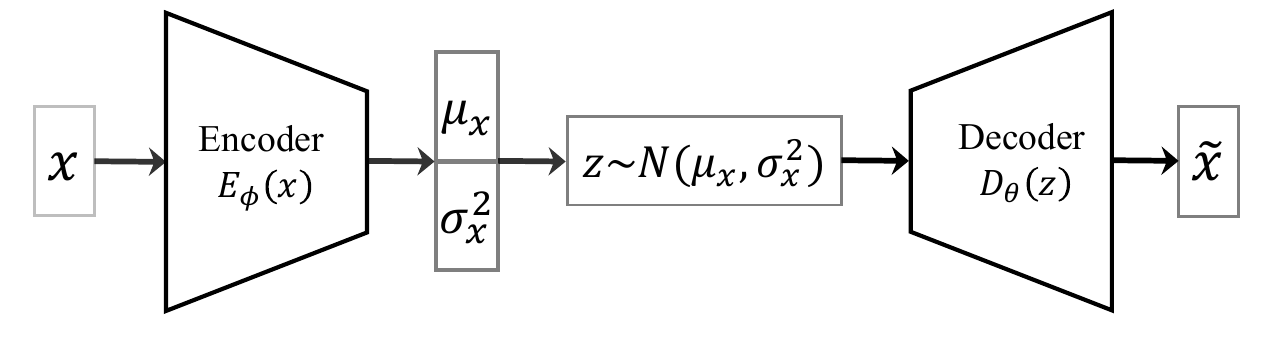}
  \caption{\label{fig:condVAE}The structure of Variational Autoencoder (VAE).}
\end{figure}

Synthetic image generation was not the primary goal of the original autoencoder, which instead focused on feature learning or dimensionality reduction. 
The VAE~\cite{ref:VAE} introduced a new generative model by combining the AE structure and the variational inference technique. 
As shown in Fig.~\ref{fig:condVAE}, the input data $\rvx$ is encoded to a Gaussian distribution $\mathcal{N}(\rvx; \evmu_x, \evsigma_x^2)$ with the Encoder $E_\phi(\rvx)$. Then, a variable $\rvz$ is sampled from the Gaussian $\rvz \sim \mathcal{N}(\evmu_x,\evsigma_x^2)$. Finally, the decoder $D_{\theta}(z)$ takes $\rvz$ to generate the reconstructed data $\tilde{\rvx}$. 
The intuition behind the VAE is to encode an input as a distribution instead of a simple point. With proper regularization of the distribution, the latent space of the VAE becomes more continuous and complete and can therefore be conveniently sampled to improve the quality of the data generation. 

For effective training of the VAE model, the following Evidence lower bound (ELBO) is optimized:
\begin{equation}
    \log p_\theta(\rvx) \geq \mathcal{L}(\theta,\phi;\rvx) = 
    -D_{KL}(q_\phi(\rvz|\vx)\|p_\theta(\rvz)) + \mathbb{E}_{q_\phi(\rvz|\rvx)} [\log p_{\theta}(\rvx|\rvz)]\,, 
\end{equation}
where $q_\phi(\rvz|\rvx)$ is a model used to approximate the intractable posterior distribution $p_\theta(\rvz|\rvx)$, $D_{KL}$ is the Kullback–Leibler divergence between two distributions to measure the divergence between two distributions, and 
$p_\theta(\rvx|\rvz)$ is the reconstruction probability of $\rvx$ given $\rvz$. In a deep VAE framework (Fig.~\ref{fig:condVAE}), $q_\phi(\rvz|\rvx)$ is modeled by an Encoder $E_\phi(\rvx)$ and $p_\theta(\rvz|\rvx)$ is modeled by a Decoder $D_\theta(\rvz)$. 
$p_\theta(\rvz)$ is set to an isotropic Gaussian to regularize the posterior distribution. Moreover, the reparameterization trick is used to facilitate the model training and the gradient estimation: $\rvz = \evmu_x + \evsigma_x\epsilon, \epsilon \sim \mathcal{N}(0,1)$.

\subsubsection{Conditional VAE (CVAE)} 
Similar to the Conditional GAN, a conditional version of the VAE model was proposed in~\cite{ref:CVAE}. 
The model is not trained to reconstruct the input $\rvx$. Instead, for the input $\rvx$, the latent variable $\rvz$ is sampled from $p_\theta(\rvz|\rvx)$, and then a different output $\rvy$ is generated according to $p_\theta(\rvy|\rvx,\rvz)$ rather than the original reconstruction $\tilde{\rvx}$. 
Based on the ELBO of the VAE, the variational lower bound of this conditional model is as follows:
\begin{equation}
    \log p_\theta(\rvy|\rvx) \geq -D_{KL} (q_\phi(\rvz|\rvx,\rvy)\|p_\theta(\rvz|\rvx)) + \mathbb{E}_{q_\phi(\rvz|\rvx,\rvy)} [\log p_\theta(\rvy|\rvx,\rvz)]\,.
\end{equation}
For a deep CVAE framework, $q_\phi(\rvz|\rvx,\rvy)$ is modeled as a recognition network, $p_\theta(\rvz|\rvx)$ is modeled as a conditional prior network, and $p_\theta(\rvy|\rvx,\rvz)$ is modeled as the generation network. All these networks can be either Multi-Layer Perceptrons (MLPs) or Convolution Neural Networks (CNNs). 

Various applications can be derived from different choices of $\rvx$ and $\rvy$. 
For example, if $\rvx$ represents the original image and $\rvy$ represents the segmentation mask, then we are dealing with an image segmentation task. 
If $\rvx$ represents an MRI image and $\rvy$ represents a CT image, then we are dealing with an MRI-to-CT translation task. 


\subsubsection{U-Net} 
\begin{figure}[h]
  \centering
  \includegraphics[width=0.7\textwidth]{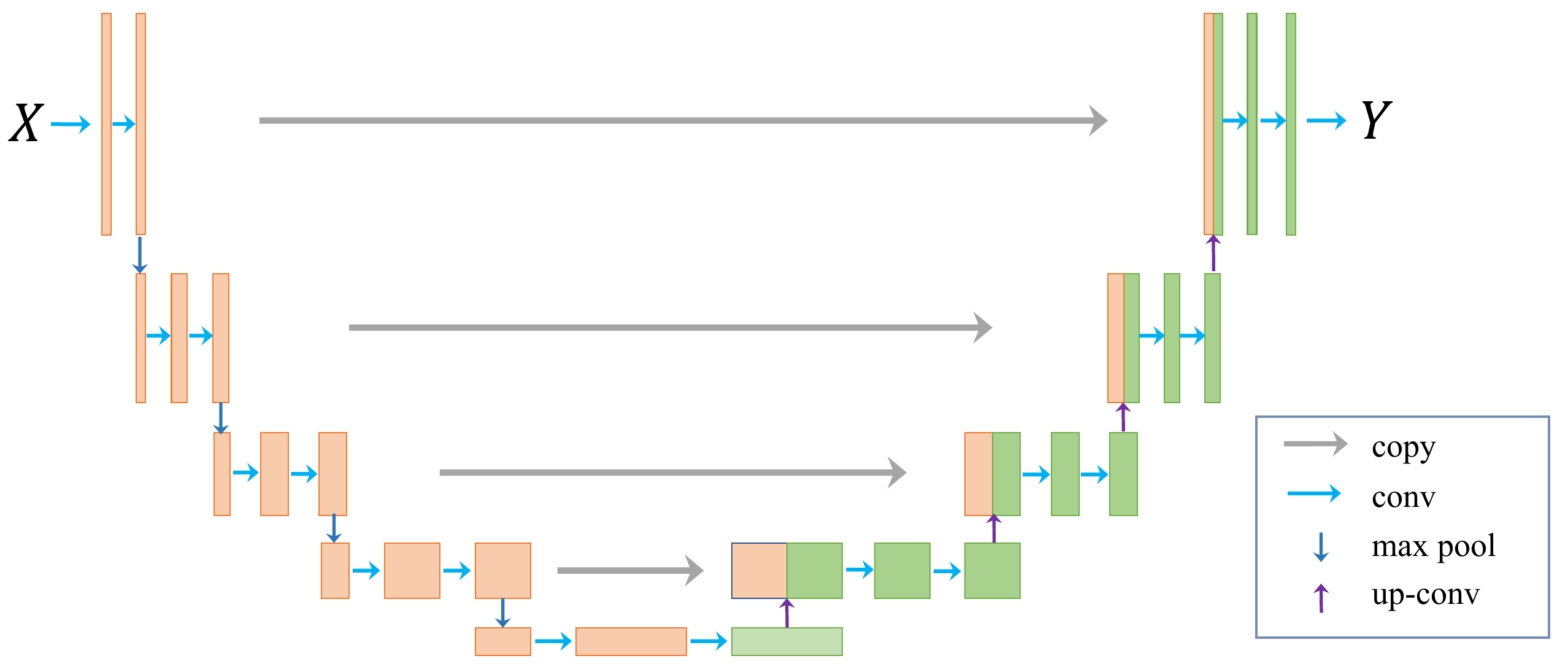}
  \caption{\label{fig:UNet}A schematic view of the U-Net architecture. Several intermediate feature maps from the encoder are copied and concatenated in the decoder.} 
\end{figure}

U-Net was first proposed in~\cite{ref:UNET} for biomedical image segmentation. It has an encoder-decoder architecture, as shown in Figure~\ref{fig:UNet}. Specifically, the encoder converts an image $X$ into a lower-dimension representation. Then, the decoder takes these representations as inputs and copies several intermediate feature maps from the encoder to generate the segmentation mask $Y$. 
Due to its excellent performance, U-Net has been widely used in medical applications besides segmentation, including cross-modality synthesis~\cite{ref:CNet}, registration~\cite{ref:VTN,ref:VoxelMorph,ref:Deform-GAN}, etc. 
For 3D medical applications, the 3D U-Net~\cite{ref:3D_UNET} was designed by using 3D convolution to process the 3D volume images.  
Moreover, the U-Net design can also be used as the main structure of other generative models, e.g., the U-Net generator in GANs.  

\subsubsection{Denoising Autoencoder (DAE)} 
DAE~\cite{ref:DAE} is an autoencoder that takes in the corrupted data at the input to reconstruct the original uncorrupted data. A noise function $T: \mathcal{X} \rightarrow \mathcal{X}$ is applied on the data $x\in \mathcal{X}$ of a standard autoencoder to construct a DAE. The objective of a DAE is to reconstruct the original data $X$ from the noisy input $T(x)$. Thus, the following optimization problem can be defined:
\begin{equation}
    \label{eq:DAE}
    \min_{\theta,\,\phi} \mathbb{E}_x\, \|x - (D_\phi \circ E_\theta \circ T )(x) \|^2 \,.
\end{equation}
In a similar way to the bottleneck design, the denoising training strategy can facilitate the learning of the inherent features of the original data. 
Besides the manually designed noise function $T$, certain medical image collection and processing devices can produce noisy and low-quality medical images, e.g., low-dose CT images. 
In this case, the idea of DAE can be applied to the medical image denoising task~\cite{ref:DAE_medical,ref:DAE_medical2}. 
%

\subsubsection{Structural Causal Models (SCMs)}
To integrate causality into deep learning methods for improved interpretability, deep structural causal models (Deep SCMs)~\cite{ref:SCMs} incorporated deep learning components into SCMs.
Given the random variable $\mathbf{x}$ and noise variable $\boldsymbol{\epsilon}$, an SCM $\mathcal{B} := (\mathbf{S}, P(\boldsymbol{\epsilon}))$ can be constructed with \emph{mechanisms} $\mathbf{S}=(f_1, \dots, f_K)$ and $x_k:=f_k(\epsilon_k; \text{pa}_k)$, where $\text{pa}_k$ is the parents of $x_k$ (\emph{direct causes}). 
Three types of deep mechanisms can be implemented: 1) invertible explicit likelihood with normalizing flow: $f_i(\epsilon_i;\text{pa}_i), p(x_i|\text{pa}_i) = p(\epsilon_i)\cdot |\text{det}_{\epsilon_i}f_i(\epsilon_i;\text{pa}_i)|^{-1}|_{\epsilon_i=f_i^{-1}(x_i;\text{pa}_i)}$; 2) amortized explicit likelihood with variational inference: $f_k(\epsilon_k;\text{pa}_k)=h_k(\mu_k;g_k(z_k;\text{pa}_k),\text{pa}_k)$, $P(\epsilon_k)=P(\mu_k)P(z_k)$, where $h_k$ is invertible and $g_k$ is non-invertible; 3) amortized implicit likelihood: use adversarial objective to train a conditional implicit model. 
With the deep mechanism, counterfactual inference can be performed in three steps: Abduction, Action, and Prediction. 
Several recent works explored the idea of interpretable models and counterfactual augmentation~\cite{ref:CRT_response,ref:cardiac_classification,ref:CounterSynth}.

\subsubsection{Denoising Diffusion Probabilistic Model (DDPM)}

Inspired by non-equilibrium thermodynamics~\cite{ref:thermodynamics}, diffusion models devise a Markov chain of diffusion steps by gradually adding noise to the data (forward) and learning how to reverse the diffusion process (reverse) so that an image can be generated from the noise. 
Recent research has demonstrated that Denoising Diffusion Probabilistic Models (DDPMs)~\cite{ref:DDPM} outperform GANs in terms of natural image synthesis~\cite{ref:beat_gans,ref:beat_gans2}. 
Given data sample $\rvx_0 \sim q(\rvx_0)$, the forward diffusion process is defined as: $q(\rvx_t|\rvx_{t-1}) = \mathcal{N} (\rvx_t; \sqrt{1-\beta_t}\rvx_{t-1}, \beta_t \rmI); q(\rvx_{1:T}|\rvx_0) = \prod_{t=1}^T q(\rvx_t|\rvx_{t-1})$, where $\beta_t$ controls the noise schedule such that $q(\rvx_T|\rvx_0) \approx \mathcal{N} (\rvx_T; \mathbf{0}, \rmI).$ 
In other words, the forward process converts a data sample into a zero-mean isotropic Gaussian. 
The reverse denoising process is defined as: $p(\rvx_T) = \mathcal{N}(\rvx_T; \mathbf{0}, \rmI);$ $p_{\theta}(\rvx_{t-1}|\rvx_{t}) = \mathcal{N} (\rvx_{t-1}; \mu_{\theta}(\rvx_t, t), \sigma_t^2\rmI)\,,$ where $\mu_{\theta}(\rvx_t, t)$ is a trainable network to estimate the mean and $\sigma_t^2$ is the variance schedule. 
The reverse process generates realistic data from Gaussian noise. 

Similar to VAE, the training of the DDPM model uses the variational upper bound to construct the following objective function: 
\begin{equation}
    L = \mathbb{E}_q \big[ \underbrace{D_{KL}(q(\rvx_T|\rvx_0) \| p(\rvx_T))}_{L_T} + \sum_{t>1} \underbrace{D_{KL} (q(\rvx_{t-1}|\rvx_t,\rvx_0) \| p_{\theta}(\rvx_{t-1}|\rvx_t))}_{L_{t-1}} - \underbrace{\log p_{\theta}(\rvx_0|\rvx_1))}_{L_0} \big] \,.
\end{equation}
To improve the training of the objective function, a simplified version of $L_{t-1}$ is proposed in~\cite{ref:DDPM}: $L_{\text{simple}} =\mathbb{E}_{\rvx_{0}\sim q(\rvx_0), \epsilon \sim \mathcal{N}(\mathbf{0},\rmI), t \sim \mathcal{U}(1,T)} [\| \epsilon - \epsilon_{\theta}(\sqrt{\bar{\alpha}_t}\rvx_0 + \sqrt{1-\bar{\alpha}_t}\epsilon, t) \|^2]$. 
Here, $\epsilon_{\theta}$ is a function approximator to predict $\epsilon$ from $\rvx_t$, and $\bar{\alpha}_t = \prod_{s=1}^t(1-\beta_s)$. $L_{\text{simple}}$ achieved better generation quality on natural image benchmarks. 

In practice, the approximation functions $\mu_{\theta}(\rvx_t, t)$ or $\epsilon_{\theta}$ can be implemented with U-net or DAE structures. Due to the good generation performance, DDPM has been adopted in several recent medical imaging papers~\cite{ref:diffusion_generation,ref:diffusion_detect_segment}. 


\subsection{Autoregressive Models}

\begin{figure}[ht]
  \centering
  \includegraphics[width=0.3\textwidth]{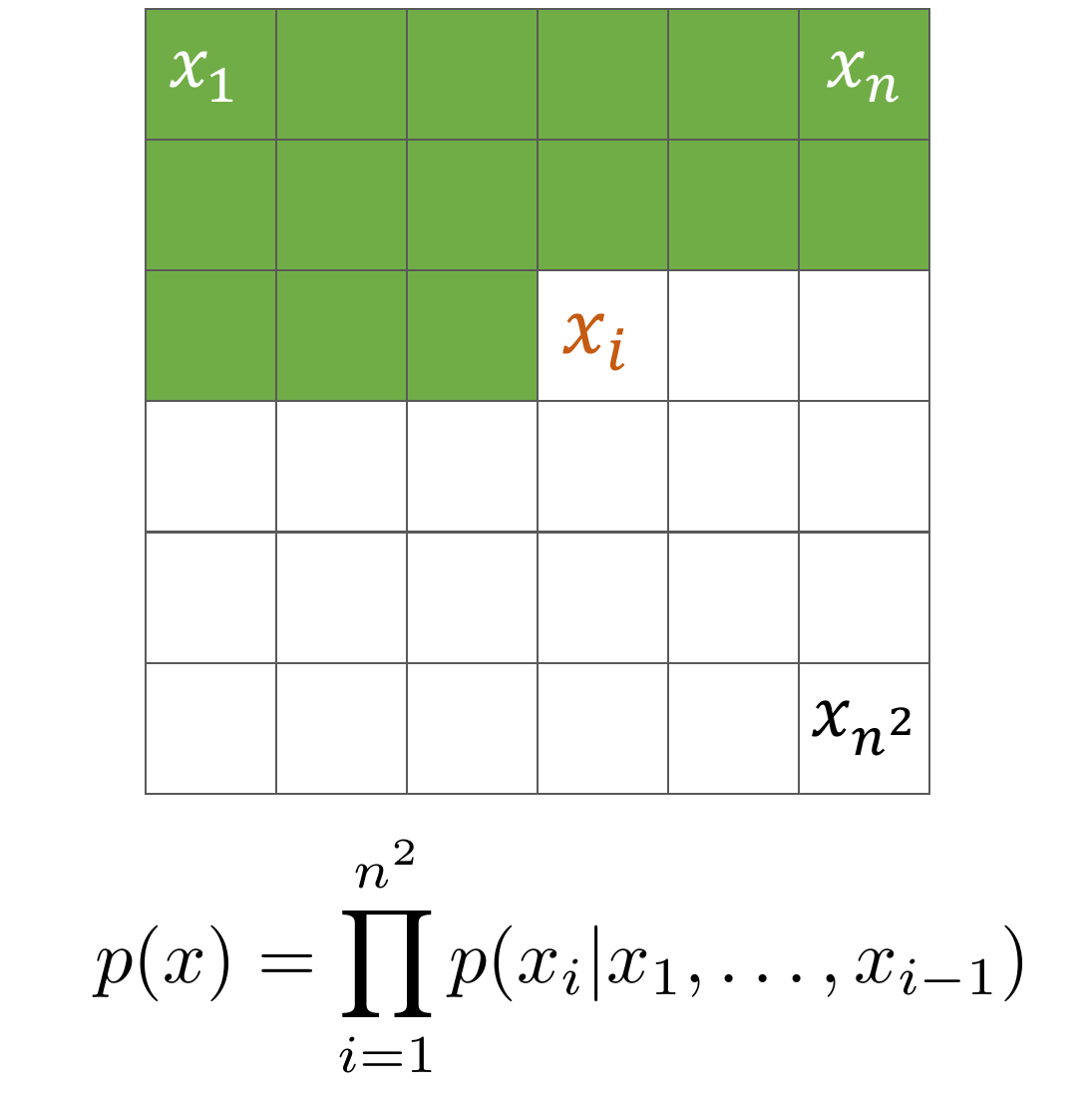}
  \caption{\label{fig:autoregressive}A schematic illustration of an autoregressive model. An image distribution is modeled as the distribution of all pixels, where each pixel is conditioned on the left and above surrounding pixels.}
\end{figure}

The hallmark problem of generative models is to model the distribution of images.
GANs apply an adversarial strategy to indirectly approach the real distribution, while VAE employs variational inference to approximate it. 
In contrast, Autoregressive models~\cite{ref:autoregressive,ref:pixelRNN} explicitly compute the joint distribution of pixels by factorizing it into products of conditional distributions. 
As shown in Fig.~\ref{fig:autoregressive}, an $n\times n$ image is represented by a sequence of pixels $x_1, x_2, \dots, x_{n^2}$ according to the raster scan order\footnote{Rater scan order: row by row and pixel by pixel within every row}. The distribution of the forthcoming pixel $x_i$ depends on the above and left surrounding pixels, i.e., $p(x_i|x_1,\dots, x_{i-1})$. 
In order to calculate the distribution of the whole image, the chain rule is applied: $p(x) = \prod_{i=1}^{n^2} p(x_i|x_1,\dots, x_{i-1})$. For a color image, each pixel $x_i$ is determined by three channels: Red, Green, and Blue. 
This is followed by the introduction of the successive dependency: $p(x_i|\rvx_{<i}) = p(x_{i,R}|\rvx_{<i}) p(x_{i,G} |\rvx_{<i}, x_{i,R}) p(x_{i,B} |\rvx_{<i}, x_{i,R}, x_{i,G})$. 

Autoregressive models calculate the tractable image distribution in an explicit and straightforward manner by maximizing the log-likelihood function. 
However, due to the factorization of multiple product terms, the direct computation is inefficient. 
Thus, a variety of autoregressive models~\cite{ref:autoregressive,ref:pixelRNN,ref:cond_pixelcnn,ref:autoregressive_transformer} have been proposed to enable fast and parallel computation. 
Below, we introduce several methods that are relevant to the synthetic generation of 3D medical volume images.


\subsubsection{PixelRNN and PixelCNN} 
In order to improve the expressive power and efficiency of early autoregressive models such as NADE~\cite{ref:NADE}, the work in \cite{ref:pixelRNN} used Recurrent Neural Networks (RNNs) to construct the PixelRNN model for large-scale natural image generation. 
PixelRNN is designed to stack multiple 2D Long Short-Term Memory (LSTM) layers. 
In each layer, there is a fixed input-to-state component and a recurrent state-to-state component to jointly compute the states and gates of an LSTM. 
Specifically, two types of LSTM layers are designed to implement the state-to-state components: Row LSTM and Diagonal BiLSTM. The Row LSTM processes the image row by row with a $k\times 1$ convolution, while the Diagonal BiLSTM skews each input row by one position and processes with a $2\times 1$ convolution. 
Residual connections~\cite{ref:resnet} are deployed to facilitate deep PixelRNN training. 

The Row LSTM and the Diagonal BiLSTM have the benefit of potentially unbounded dependency range to model the global pixel correlations. However, they introduce large computation costs. Therefore, another variant (called PixelCNN) using standard convolution layers along with masks was proposed to enable a fast and parallel training process. 
The speedup of PixelCNN, however, degrades the generation performance when compared with the PixelRNN model.

\subsubsection{Autoregressive Transformer}
Compared with RNNs and CNNs, Transformers~\cite{ref:transformer} have shown superior performance in modeling the long-range interactions within sequences. 
This inspired the Autoregressive Transformer proposed in \cite{ref:autoregressive_transformer}. 
Specifically, a Vector Quantised GAN (VQGAN) was designed to learn a representative codebook from the patches of the training images. 
The composition of the codes in an image was then modeled using an autoregressive transformer. 
Finally, image generation was formulated as the autoregressive prediction of the next image patch, similar to PixelRNN. 
Conditional synthesis was also achieved in \cite{ref:autoregressive_transformer} by enforcing the conditional information in the autoregressive transformer.

\subsection{Advantages and Disadvantages of Different Generative Models}

GANs have become a preferred choice for medical image generation, offering multiple advantages such as (1) \emph{High-Quality Output}: They are capable of generating highly realistic synthetic images, as evidenced by their top-tier FID scores~\cite{ref:FID}, which can be used to augment limited medical image datasets. (2) \emph{Flexibility}: Their versatility allows them to be applied to a variety of tasks, including image augmentation, segmentation, super-resolution, and denoising. (3) \emph{Unpaired Cross-Modality Synthesis}: With the use of CycleGAN~\cite{ref:un_image_to_image}, GANs can convert between different imaging modalities like MRI to CT without requiring paired data, thereby reducing data collection burdens. Despite these benefits, GANs also have notable drawbacks, including (1) \emph{Training Challenges}: They are notoriously difficult to train due to issues like training instability~\cite{ref:gans_instability1,ref:gans_instability2} and mode collapse~\cite{ref:gans_mode1,ref:gans_mode2}, which arise from the oscillatory behavior of the two competing networks and vanishing gradients. (2) \emph{Uninterpretable Latent Space}: The lack of an explicit latent space makes it challenging to interpret the patterns in the generated images, limiting their applicability in clinical decision-making.

While AEs may not match GANs in terms of generation quality, they offer their own set of advantages and can address some of GANs' limitations in specific contexts. These advantages include (1) \emph{Latent Space Representation}: AEs feature an explicit latent space that facilitates representation learning and feature interpretation. (2) \emph{Stable Training}: The use of reconstruction loss and regularization contributes to a more stable training process. (3) \emph{Reconstruction Capability}: AEs are naturally suited for tasks like medical image denoising due to their reconstruction properties. However, AEs also come with their own set of drawbacks, such as (1) \emph{Blurred Output}: The averaging effect in the latent space often results in less sharp images, particularly in models like VAEs. (2) \emph{Limited Expressivity}: The constrained dimensionality of the latent space can limit the model's ability to generate high-dimensional data distributions.

Autoregressive models may be less common than GANs and AEs, but they excel as sequential generators. Their strengths include (1) \emph{Explicit Distribution}: These models explicitly learn the joint distribution of pixels, enhancing both transparency and interoperability. (2) \emph{Sequential Processing}: They are particularly effective for generating medical time-series data, such as 4D spatio-temporal anatomy~\cite{ref:R3_p6}, spatio-temporal registration~\cite{ref:R3_p3}, or Time-of-Flight~\cite{ref:TOF-MRA}. However, their sequential nature also introduces some drawbacks, including (1) \emph{Inefficient Parallelization}: The sequential architecture creates dependencies between pixels, making the training process slower. (2) \emph{Absence of Latent Space}: Unlike GANs and AEs, autoregressive models do not have a latent feature space.

In the field of medical imaging, the choice of model is contingent on both the specific application and its requirements. For instance, GANs are generally the preferred choice when high-quality generation or cross-modality capabilities are needed. On the other hand, AEs are more suitable when the focus is on reconstruction or interpretability. For tasks involving sequential imaging data, Autoregressive models naturally come to the forefront. To aid readers in making an informed model selection, Section~\ref{sec:application} provides a comprehensive list of models used in various papers and applications.

\subsection{Challenges of 3D Generative Models}
The aforementioned generative models have been applied to various two-dimensional (2D) tasks, including natural image generation~\cite{ref:stylegan,ref:stylegan2}, image style transfer~\cite{ref:image_to_image,ref:un_image_to_image}, text-to-image generation~\cite{ref:cond_gans_text}, medical slice synthesis~\cite{ref:gans_survey1}, 2D cross-modality synthesis~\cite{ref:GLA-GAN,ref:MRI-Trans-GAN}, etc. 
However, as a result of the intrinsic 3D nature of medical volume images, there are practical challenges that are beyond the design and training of 2D generative models.

\textbf{Insufficient 3D medical volumes for training.} 
A 2D generative model relies on a large number of readily accessible images (e.g., natural images, slices of medical volume images) in order to model real data distributions. 
For example, the widely-used ImageNet~\cite{ref:ImageNet} dataset contains 1.28 million natural images. In the medical field, a big dataset for training can be created by compiling all the 2D slices of a 3D volume. 
Comparatively, 3D medical datasets are several orders of magnitude smaller. 
3DSeg-8~\cite{ref:med3d} consists of eight medical datasets, with each dataset containing only tens to hundreds of volume images. 
For the training of generative models, \cite{ref:3D_a_WGANGP} utilized only 991 brain T1-weighted image volumes from ADNI~\cite{ref:ADNI}. 

\textbf{3D model parameters.} 
3D generative models usually utilize 3D convolution architectures, such as 3D-U-Net~\cite{ref:3D_UNET}. 
While 3D convolution architectures are capable of extracting feature representations pertaining to 3D anatomical structures, they also require significantly more parameters compared to their 2D counterparts. 
Consequently, a number of problems will occur: \textbf{(1)} the deployment and training of 3D models are slow, \textbf{(2)} 3D models consume a lot of GPU memory, which limits the training batch size, \textbf{(3)} the lack of sufficient 3D volume images for training large 3D models leads to overfitting.  

\textbf{Ultra-high dimension of the 3D volumes.} 
In contrast to 2D images, 3D volume images have a very high dimension, which makes modeling the 3D volume distribution extremely difficult. 
In practice, it is infeasible to design a 3D generative model that captures the 3D volume distribution exactly. 
As a result, approximation techniques such as variational inference or distribution factorization need to be adopted. 
The resolution of synthesized images is usually limited to small values, such as $64\times 64\times 64$ or $128\times 128\times 128$, to reduce the data dimension. 
In contrast, 2D image synthesis can use resolutions of up to $1024\times 1024$. 

\section{3D Brain and Heart Applications of Generative Models}
\label{sec:application}

In each of the following subsections, we describe the application background, elaborate on the methods for brain and heart applications, and provide a summary table showing the investigated organ, the model, the datasets, and the performance metrics.

\subsection{Unconditional Synthesis}

\begin{table}[ht]
\centering
\setlength{\tabcolsep}{3pt}
\caption{\label{tab:uncond_synthesis}Summary of publications on 3D Unconditional Synthesis related to the brain or heart.}
\resizebox{1.0\textwidth}{!}{
\begin{tabular}{l|l|l|l|l}\hline
Publication \small{(Year)}                        & Organ & Model & Dataset             &   Metrics       \\\hline
\small{3D-$\alpha$-WGAN-GP~\cite{ref:3D_a_WGANGP} (2019)} & Brain & $\alpha$-GAN~\cite{ref:a_gan} & \small{ADNI~\cite{ref:ADNI}, BRATS 2018~\cite{ref:brats18}}  &  MMD, MS-SSIM \\
 & & WGAN-GP~\cite{ref:wgan_gp} & ATLAS~\cite{ref:ATLAS} & \\ \hline
SlabGAN~\cite{ref:SlabGAN} \small{(2019)} & Brain & \small{ProgressiveGAN~\cite{ref:progressiveGAN}} & fastMRI~\cite{ref:fastMRI} & FID \\ \hline
{\"O}zbey \emph{et al}.~\cite{ref:Ozbey} \small{(2020)} & Brain & GAN~\cite{ref:gans} & IXI~\cite{ref:IXI} & PSNR, FID \\\hline
2D Slice VAE~\cite{ref:slice_vae} \small{(2020)}  & Brain & VAE~\cite{ref:VAE} & HCP~\cite{ref:HCP}  &  \small{RAS, MMD, MS-SSIM} \\\hline
3D StyleGAN~\cite{ref:3d_stylegan} \small{(2021)} & Brain & StyleGAN2~\cite{ref:stylegan2} & \small{ADNI~\cite{ref:ADNI}, ABIDE~\cite{ref:ABIDE}, HABS~\cite{ref:HABS}} & \small{bMMD$^2$, MS-SSIM, FID} \\
               &       &           & \small{OASIS~\cite{ref:OASIS}, MCIC~\cite{ref:MCIC}, PPMI~\cite{ref:PPMI}}   &\\
               &       &           & \small{ADHD~\cite{ref:ADHD200}, Harvard, GSP~\cite{ref:harwardGSP}} & \\\hline
DCR-$\alpha$-GAN~\cite{ref:DCR-a-GAN} \small{(2021)}  & Brain & $\alpha$-GAN~\cite{ref:a_gan} & ADNI~\cite{ref:ADNI} & \small{MMD, MS-SSIM, IoU} \\\hline
\small{Shape+Texture GAN~\cite{ref:shape_texture_gan} (2021)} & Brain & WGAN-GP~\cite{ref:wgan_gp} & HCP~\cite{ref:HCP}, ADNI~\cite{ref:ADNI} & MMD, MS-SSIM \\\hline
\small{Split\&Shuffle-GAN~\cite{ref:split_shuffle} (2022)}& Brain & StyleGAN2~\cite{ref:stylegan2}  & COCA~\cite{ref:COCA}, ADNI~\cite{ref:ADNI} &  FID   \\
 & Heart &   &  &    \\\hline
HA-GAN~\cite{ref:HA_GAN} \small{(2022)} & Brain & GAN~\cite{ref:gans}, AE~\cite{ref:AE1} & Harvard GSP~\cite{ref:harwardGSP} &  FID, MMD, IS   \\\hline
DDM~\cite{ref:diffusion_generation} \small{(2022)} & Heart & DDPM~\cite{ref:DDPM}  & ACDC~\cite{ref:ACDC} & \small{PSNR, NMSE, Dice} \\ \hline

\end{tabular}  
}
\end{table}

The most straightforward application of generative models for 3D medical imaging is unconditional synthesis.
By accurately representing the real data distributions of 3D medical volumes, synthesized images of high quality can be produced for downstream tasks such as tumor prediction~\cite{ref:tumor_prediction} and Alzheimer's disease diagnosis~\cite{ref:3DCapsNet,ref:Lin_crossmodality}.
These simulated images can help mitigate two major challenges in data collection: (1) data deficit, which arises from the need for specialized annotation, privacy concerns, and financial constraints, and (2) data imbalance in terms of attributes like health status, age, race, and skin color. 
Compared to 2D medical imaging, 3D medical imaging places a greater emphasis on the utility of synthesized images. 
The number of 3D volume images in existing medical datasets ranges from hundreds~\cite{ref:med3d} to thousands~\cite{ref:3d_stylegan}, which is not enough to train a reliable deep diagnosis model for clinical use. 
With the rapid development of generative models for natural images, a number of successful architectures have been modified for 3D medical volume synthesis, e.g., VAE~\cite{ref:VAE}, Progressive GAN~\cite{ref:progressiveGAN}, StyleGAN~\cite{ref:stylegan2}, $\alpha$-GAN, etc. 
There are a number of publications that describe unconditional synthesis, as shown in Table~\ref{tab:uncond_synthesis}. 


\subsubsection{\textbf{AE and VAE}}
AE and VAE structures are commonly used in medical generative models, either alone or in combination with GANs. 
Based on a combined two-dimensional slice-level VAE and slice correlation model,  
2D Slice VAE~\cite{ref:slice_vae} generated 3D MRI brain volumes. 
The VAE was simply trained on MRI brain slices to form a latent space, from which the latent code was sampled for slice generation. To ensure the completeness and consistency of the slices in the generated volumes, the means and covariances of each dimension of the latent code were separately estimated from the training slices. 
Moreover, \cite{ref:slice_vae} proposed a Realistic Atlas Score (RAS) based on the affine-registered segmentations of the generated volumes as an evaluation metric. 
As a generalization of the VAE model, DDPM~\cite{ref:DDPM} was used in DDM~\cite{ref:diffusion_generation} for the task of 4-Dimensional (3D+t) temporal cardiac MRI generation. A Diffusion module was trained to estimate the latent code $c$, which was then leveraged by the Deformation module to generate deformed images with the spatial transformation layer~\cite{ref:STN}. 
3D-$\alpha$-WGAN-GP~\cite{ref:3D_a_WGANGP} adapted the $\alpha$-GAN structure, which combines both the VAE and GAN into a single model to generate 3D brain MRI volumes from random vectors. The architecture consists of a Generator, a Discriminator, an Encoder, and a Code Discriminator. To improve the training stability, the model was trained with the WGAN-GP loss~\cite{ref:wgan_gp} and the reconstruction loss. The integration of VAE and GAN simultaneously addressed the mode collapse and image blurriness problems. 
Similarly, the $\alpha$-GAN structure was also employed in DCR-$\alpha$-GAN~\cite{ref:DCR-a-GAN}, which integrated a Refiner Network with four ResNet blocks after each generator to refine the generation process and produce more realistic images. 

\subsubsection{\textbf{GANs}}
In GANs, the progressive growing strategy~\cite{ref:progressiveGAN} has been verified to improve the quality and stability of image generation. 
SlabGAN~\cite{ref:SlabGAN} directly extended the 2D ProgressiveGAN~\cite{ref:progressiveGAN} for the 3D generation of MRI brain images. Specifically, SlabGAN replaced the original $4\times 4$ constant with $4\times 4 \times 4$ to finally generate volumes of $256\times 256\times 16$ resolution. 
HA-GAN~\cite{ref:HA_GAN} designed a hierarchical generation structure containing two branches: the full volume low-resolution generation branch and the sub-volume high-resolution generation branch. Each branch consists of a GAN network and an AE network with a shared decoder/generator. The two branches also share the low-resolution generation block to facilitate progressive generation. 
{\"O}zbey \emph{et al}.~\cite{ref:Ozbey} progressively generated 3D volumes with three sequential GAN models. An Axial GAN $G_A$ generated MRI volumes by concatenating cross-sections~\cite{ref:cross_section} in the axial view. Taking the generated volumes as inputs, a Coronal GAN $G_C$ was trained to enhance the quality of the coronal view. 
Similarly, a Sagittal GAN $G_s$ enhanced the previously generated volumes in the sagittal direction. 
The Shape+Texture GAN~\cite{ref:shape_texture_gan} incorporated the geometric deformation technique~\cite{ref:deformation} in two GANs: Shape Network and Texture Network. Shape Network was trained to generate the 3D deformation field from a random vector. 
To obtain synthetic images, the generated 3D fields were applied to the Montreal Neurological Institute (MNI) brain template to warp it. 
Both deformed and real images were utilized to train the Texture Network composed of a U-Net~\cite{ref:UNET} and a patch-GAN~\cite{ref:patchGAN} discriminator.

\subsubsection{\textbf{StyleGAN}}
As a result of the style-based generation technique~\cite{ref:stylegan,ref:stylegan2}, StyleGAN~\cite{ref:stylegan} no longer requires the progressive generation strategy~\cite{ref:progressiveGAN}. 
3D StyleGAN~\cite{ref:3d_stylegan} extended the state-of-the-art 2D StyleGAN2~\cite{ref:stylegan2} architecture for 3D medical images generation. 
Specifically, the mapping network for generating each layer's style vectors remained unchanged from 2D to 3D. The generator was lifted to 3D with 3D convolution operations and the $4\times 4$ constant input was replaced with $5\times 6 \times 7$ to fit the 3D brain volumes. The discriminator was also lifted with 3D convolutions. A total of $7,392$ image volumes were collected from eight datasets to train the model. 
The Split\&Shuffle-GAN~\cite{ref:split_shuffle} addressed the training challenges of the 3D StyleGAN from the perspective of the training strategy and network architecture. 
First, a 2D CNN was trained, and then five strategies were proposed to inflate the 2D weights to 3D. The Channel Split\&Shuffle modules were respectively devised for the discriminator and the generator to significantly reduce the number of parameters. Both the brain and heart synthetic volume images were generated in \cite{ref:split_shuffle}. 

\subsubsection{\textbf{Quality Assessment}}
In evaluating the quality of 3D synthetic volume images, it is important to consider the 3D anatomical structure, which is not evident in a 2D image. 
As a result, \cite{ref:3d_evaluation} assessed various image quality metrics on MRI images generated by GANs, and proposed a Deep Quality Assessment (QA) model that was trained according to the human ratings of MRI images. 
Among all, Fréchet Inception Distance (FID)~\cite{ref:FID}, Naturalness Image Quality Evaluator (NIQE)~\cite{ref:NIQE}, and Maximum Mean Discrepancy (MMD)~\cite{ref:MMD} showed good sensitivity to low-quality images. Deep Quality Assessment managed to distinguish between high-quality images exhibiting subtle variations.


\subsection{Classification}

\begin{table}[ht]
\centering
\caption{\label{tab:classification}Summary of publications on 3D Classification related to the brain or heart.}
\resizebox{1.0\textwidth}{!}{
\begin{tabular}{l|l|l|l|l}\hline
Publication \small{(Year)}     & Organ &Model     & Dataset         & Metrics       \\\hline
SCGANs~\cite{ref:SCGANs} \small{(2017)}     & Heart & GAN~\cite{ref:gans}  & UK Biobank~\cite{ref:UKbiobank} & ACC, Pre, Rec \\\hline
Biffi \emph{et al}.~\cite{ref:cardiac_classification} \small{(2018)} & Heart & VAE~\cite{ref:VAE} & Multi-centre~\cite{ref:cardiac_classification}, ACDC~\cite{ref:ACDC} & ACC  \\\hline
3D-CapsNet~\cite{ref:3DCapsNet} \small{(2019)} & Brain & AE~\cite{ref:AE1} & ADNI~\cite{ref:ADNI} & ACC, ROC  \\\hline
3DPixelCNN~\cite{ref:3DPixelCNN} \small{(2019)} & Brain & PixelCNN~\cite{ref:pixelRNN} & DWI~\cite{ref:3DPixelCNN}, GM~\cite{ref:3DPixelCNN} & Dice, ACC, MSE \\\hline
Puyol-Ant{\'o}n \emph{et al}.~\cite{ref:CRT_response} \small{(2020)} & Heart & VAE~\cite{ref:VAE} & Biobank~\cite{ref:UKbiobank} & \small{Balanced ACC, Sen, Spe} \\\hline

\multicolumn{5}{l}{ACC: accuracy. Pre: precision. Rec: recall. Sen: sensitivity. Spe: specificity.}\\

\end{tabular}   
}
\end{table}

The classification of medical images has a multitude of applications, including Alzheimer’s disease detection~\cite{ref:3DCapsNet}, cardiovascular diseases diagnosis~\cite{ref:cardiac_classification}, age/sex prediction~\cite{ref:CounterSynth,ref:3DPixelCNN}, etc. 
For the highest level of accuracy, these applications rely on anatomical structures within the 3D medical volume. 
To apply deep learning techniques for 3D medical volume classification, 3D CNNs are commonly used ~\cite{ref:3DCNN,ref:3d_survey,ref:med3d,ref:3d_segmentation_survey}. 
Two problems arise when 3D CNNs are used for 3D volume classification: insufficient training data and lack of explainability. 
Synthesizing volumes (either unconditionally or conditionally) to augment the training data can address the issue of insufficient data. 

Clinicians are interested not just in the classification outcomes but also in explanations that pinpoint specific regions of the image that support these results~\cite{ref:explainable,ref:trustworthy}.
This can be accomplished by interpreting the classifier or learning meaningful latent spaces~\cite{ref:CRT_response}. 
Models that offer explanations will enhance trust among clinicians and expedite the transition of deep learning algorithms into practical diagnostic tools in a clinical setting. 
 
A list of generative models for the 3D classification task is shown in Table~\ref{tab:classification}.


SCGANs~\cite{ref:SCGANs} tackled the problem of classifying missing apical slices (MAS) and missing basal slices (MBS). A Semi-Coupled GANs was designed to include two generators, one for positive and one for negative data generation. A shared discriminator was designed to perform two classification tasks: real/fake and positive/negative. After GANs training, the MAS/MBS classifier was trained on both generated and real data. 
Puyol-Ant{\'o}n \emph{et al}.~\cite{ref:CRT_response} performed interpretable Cardiac Resynchronisation Therapy response prediction using a VAE and several classifiers. The volume data was constructed by temporally sampling $T=25$ points to extract 2D echocardiography images. VAE processed the slice-level segmentations and concatenated all the $T$ latent codes. The primary classifier and the multiple concept classifiers used the latent codes for prediction and interpretation, respectively.  
In order to accurately detect Alzheimer's disease in its early stages with small-scale datasets, 3D-CapsNet~\cite{ref:3DCapsNet} was designed to perform Content-Based Image Retrieval (CBIR). 
The model consisted of a pre-trained sparse Autoencoder, a 3D Capsule Network~\cite{ref:capsule}, and a 3D Convolutional Neural Network~\cite{ref:3DCNN}. 
3DPixelCNN~\cite{ref:3DPixelCNN} directly extended the PixelCNN~\cite{ref:pixelRNN} model to handle 3D brain MRI volumes. Moreover, by applying SpatialDropout~\cite{ref:SpatialDropout}, the 3DPixelCNN became a Bayesian model capable of performing uncertainty estimating~\cite{ref:bayesian_dropout}. 
The following three semi-supervised tasks were evaluated: semantic segmentation of lesions, age regression, and sex classification. 
Biffi \emph{et al}.~\cite{ref:cardiac_classification} proposed an interpretable model that can be used to automatically classify cardiac diseases caused by structural remodeling, e.g., hypertrophic cardiomyopathy. The segmentation masks from end-diastolic and end-systolic phases were input to a VAE model to learn meaningful latent distributions. Mean vectors of the latent distributions were used to train an MLP classifier. The partial derivative of the class labels w.r.t the mean vector was iteratively updated so as to generate interpretable visualization results.

\subsection{Conditional Synthesis}
\label{sec:conditional_synthesis}

\begin{table}[ht]
\centering
\caption{\label{tab:condition_synthesis}Summary of publications on 3D Conditional Synthesis related to the brain or heart.}
\resizebox{1.0\textwidth}{!}{
\begin{tabular}{l|l|l|l|l}\hline
Publication \small{(Year)}                        & Organ & Model & Dataset             &   Metrics       \\\hline
PAN~\emph{et al}.~\cite{ref:PAN} \small{(2018)} & Brain & CycleGAN~\cite{ref:un_image_to_image} & ADNI~\cite{ref:ADNI} & \small{ACC, Sen, Spe, PSNR} \\
& & & & \small{F1-score, AUC, MCC} \\\hline 
3D cGAN~\cite{ref:3DcGAN} \small{(2018)} & Brain & cGAN~\cite{ref:cond_gans} & BRATS~\cite{ref:brats18} & PSNR, NMSE, Dice \\\hline 
Liao~\emph{et al}.~\cite{ref:Liao_crossmodality} \small{(2018)} & Heart & GAN~\cite{ref:gans} & Own~\cite{ref:Liao_crossmodality} & Dice, ASSD \\\hline 

Joyce \emph{et al}.~\cite{ref:Joyce} \small{(2019)} & Heart & VAE~\cite{ref:VAE} & ACDC~\cite{ref:ACDC} & Dice \\ \hline  
Ea-GANs~\cite{ref:Ea-GANs} \small{(2019)} & Brain & GAN~\cite{ref:gans} & IXI~\cite{ref:IXI}, BRATS~\cite{ref:brats18} & PSNR, NMSE, SSIM \\\hline 
TPSDicyc~\cite{ref:R1_p5_crossD_synthesis} \small{(2019)} & Brain & CycleGAN~\cite{ref:un_image_to_image} & IXI~\cite{ref:IXI} & MSE, PSNR, SSIM \\ \hline 

Zhang \emph{et al}.~\cite{ref:Zhang_recons} \small{(2020)} & Brain & ESRGAN~\cite{ref:ESRGAN} & IXI~\cite{ref:IXI} & PSNR, SSIM \\\hline 
dEa-SA-GAN~\cite{ref:dEa-SA-GAN} \small{(2020)} & Brain & GAN~\cite{ref:gans}, AE~\cite{ref:AE1} & BRATS~\cite{ref:brats18}, SISS2015~\cite{ref:ISLES} & PSNR, NMSE, SSIM \\\hline 
3D-RevGAN~\cite{ref:3D-RevGAN} \small{(2020)} & Brain & RevGAN~\cite{ref:RevGAN} & ADNI~\cite{ref:ADNI} & \small{RMSE, PSNR, SSIM} \\ 
& & & & \small{ACC, Sen, Spe, AUC} \\\hline
QSMGAN~\cite{ref:QSMGAN} \small{(2020)} & Brain & WGAN-GP~\cite{ref:wgan_gp} & Own~\cite{ref:QSMGAN}  & L1, PSNR, NMSW \\
& & & & HFEN, SSIM \\\hline 
XCAT-GAN~\cite{ref:XCAT-GAN} \small{(2020)} & Heart & cGAN~\cite{ref:cond_gans} & \small{ACDC~\cite{ref:ACDC}, SCD~\cite{ref:SCD}, York~\cite{ref:York}} & DSC, HD \\\hline
Mcmt-gan~\cite{ref:Mcmt-gan} \small{(2020)} & Brain & GAN~\cite{ref:gans} & IXI~\cite{ref:IXI}, NAMIC~\cite{ref:NAMIC} & PSNR, SSIM, Dice \\\hline 
$\mathbb{C}$Net~\cite{ref:CNet} \small{(2020)} & Heart & U-Net~\cite{ref:UNET} & Own~\cite{ref:CNet} & MSE, SSIM \\\hline 

Lin~\emph{et al}.~\cite{ref:Lin_crossmodality} \small{(2021)} & Brain & RevGAN~\cite{ref:RevGAN} & ADNI~\cite{ref:ADNI} & \small{PSNR, SSIM, Accuracy} \\ 
& & & & \small{Sen, Spe, AUC} \\\hline
CAE-ACGAN~\cite{ref:CAE-ACGAN} \small{(2021)} & Brain & \small{CVAE~\cite{ref:CVAE}, ACGAN~\cite{ref:ACGAN}} & Own~\cite{ref:CAE-ACGAN} & PSNR, SSIM, MAE \\\hline
CACR-Net~\cite{ref:CACR-Net} \small{(2021)} & Brain & GAN~\cite{ref:gans} & BRATS~\cite{ref:brats18} & NMSE, SSIM, PSNR \\\hline
DiCyc~\cite{ref:R1_p4_brain_crossD} \small{(2021)} & Brain & CycleGAN~\cite{ref:un_image_to_image} & IXI~\cite{ref:IXI} & MSE, PSNR, SSIM \\ \hline 

ProvoGAN~\cite{ref:ProvoGAN} \small{(2022)} & Brain & GAN~\cite{ref:gans} & In vivo Brain~\cite{ref:ProvoGAN}, IXI~\cite{ref:IXI} & PSNR, SSIM \\\hline 
ResViT~\cite{ref:ResViT} \small{(2022)} & Brain & AE~\cite{ref:AE1} & IXI~\cite{ref:IXI}, BRATS~\cite{ref:brats18} & PSNR, SSIM \\\hline 
\small{Subramaniam~\emph{et al}.~\cite{ref:Subramaniam} (2022)} & Brain & WGAN~\cite{ref:wgan_gp} & \small{PEGASUS~\cite{ref:PEGASUS}, 1000Plus~\cite{ref:1000Plus}} & FID, PRD, DSC, bAVD \\\hline 

CounterSynth~\cite{ref:CounterSynth} \small{(2022)} & Brain & StarGAN~\cite{ref:StarGAN} & UK Biobank~\cite{ref:UKbiobank} & ACC, MAE, FID \\\hline 

HDL~\cite{ref:R1_p1_cardiac} \small{(2022)} & Heart & Pixel-to-pixel~\cite{ref:image_to_image} & 3Dir MVM~\cite{ref:R1_p1_dataset} & \small{MSE, PSNR, SSIM} \\
\hline

Qiao~\emph{et al.}~\cite{ref:R3_p4} \small{(2022)} & Heart & CycleGAN~\cite{ref:un_image_to_image} & UK Biobank~\cite{ref:UKbiobank} & \small{DICE, HD, ASSD} \\
\hline

\multicolumn{5}{l}{Own: The dataset was originally collected and processed by the authors of the paper.}\\
\multicolumn{5}{l}{ACC: accuracy. Sen: sensitivity. Spe: specificity. PRD: precision and recall of distributions. }\\
\multicolumn{5}{l}{bAVD: balanced average Hausdorff distance. DSC: dice similarity coefficient.}\\
\multicolumn{5}{l}{ASSD: average symmetric surface distance. MCC: Matthews correlation coefficient}

\end{tabular}   
}
\end{table}

In the case of unconditional synthesis, high-quality 3D medical volumes can be generated solely from random input variables. This limits its applications to a broader range of tasks such as class-conditional generation~\cite{ref:CounterSynth,ref:XCAT-GAN}, reconstruction~\cite{ref:CNet,ref:Zhang_recons}, and cross-modality synthesis~\cite{ref:PAN,ref:3D_c-GANs,ref:dEa-SA-GAN,ref:Liao_crossmodality,ref:GLA-GAN,ref:MRI-Trans-GAN}. 
These tasks can be modeled into the conditional synthesis framework, where extra information can be conditioned for medical volume generation (Fig.~\ref{fig:condition}(b)). 
Depending on the conditional information, a vector or an imaging modality can be used, resulting in models like Conditional GAN~\cite{ref:cond_gans} or image-to-image translation~\cite{ref:image_to_image}. 
The formulation of the task and the design of the model depend heavily on how the conditional information is incorporated, which will be elaborated below. 
A list of relevant models for conditional synthesis is provided in Table~\ref{tab:condition_synthesis}. 


\subsubsection{\textbf{Class / Attribute Condition.}} 
CounterSynth~\cite{ref:CounterSynth} proposed a counterfactual augmentation approach to generate biological-plausible 3D brain volumes conditioned on demographic attributes. 
Specifically, the U-Net generator is conditioned on the real brain image and counterfactual attributes (e.g., age, sex) to produce the deformation fields, which are then applied to the real brain image to obtain the counterfactual brain image. The discriminator consists of a real/counterfactual classifier and an attribute classifier. 
By conditioning on label maps instead of vectors, XCAT-GAN~\cite{ref:XCAT-GAN} can synthesize high-fidelity cardiac MRI images. 
First, a U-Net~\cite{ref:UNET} was trained to obtain the 4-class or 8-class segmentation masks. The real images and masks were then fed as inputs to conditional GANs for image synthesis. 
The downstream cardiac cavity segmentation task was used to evaluate the generated data after the GANs training. 
In order to produce 3D cardiac synthetic MRI images, Joyce \emph{et al}.~\cite{ref:Joyce} conditioned the synthesis on a learned anatomical model. 
The samples from the anatomical model and random Gaussian noises were input to the sequential transform and render modules to generate synthetic volumes. All the model parameters were trained through self-supervision with a reconstruction error. 
Xia~\emph{et al.}~\cite{ref:R3_p2_brain} conditioned on difference age vectors to synthesize age-progressed 2D slices, which had the potential for 3D volumes. 
Qiao~\emph{et al.}~\cite{ref:R3_p4} considered both age and gender in a conditional generative model to generate 3D anatomy of the ageing heart. The model consisting of a cycle-consistent reconstruction and a self-reconstruction module was applied to cross-sectional and longitudinal datasets. 

\subsubsection{\textbf{Cross-modality Synthesis.}} 
Among the early works, 3D cGAN~\cite{ref:3DcGAN} extended the 2D image-to-image translation GANs model~\cite{ref:image_to_image} to synthesize the 3D FLuid-Attenuated Inversion Recovery (FLAIR) from 3D T1 MRI images. Local adaptive fusion was applied to improve the FLAIR details. 
dEa-SA-GAN~\cite{ref:dEa-SA-GAN} proposed sample adaptive GAN models to explore both the global space mapping and the local space mapping. The whole model consists of two paths. The baseline path is a standard 3D GAN used to learn global mapping. The sample-adaptive path uses the neighborhood relationships between the training samples to force local sample-specific learning. Two synthesis tasks were conducted in dEa-SA-GAN: T1$\rightarrow$FLAIR and T1$\rightarrow$T2.  
CACR-Net~\cite{ref:CACR-Net} designed a confidence-guided aggregation and cross-modality refinement network for the multi-modality MRI synthesis task. The Confidence-Guided Aggregation Module uses two U-Net generators to process two different input modalities. The two outputs are then aggregated and input to the Cross-Modality Refinement Module for further improvement. Three synthesis tasks were conducted: T1+T2$\rightarrow$FLAIR, T1+FLAIR$\rightarrow$T2, and T2+FLAIR$\rightarrow$T1. 
For CAE-ACGAN~\cite{ref:CAE-ACGAN}, only a single CT image was used to generate multi-contrast MRI images (i.e., fat, R2, and water). The Conditional Variational AutoEncoder (CVAE)~\cite{ref:CVAE} and ACGAN~\cite{ref:ACGAN} architectures were developed to take advantage of the capabilities of both VAE and GAN. The structure of CAE-ACGAN consists of an Encoder, a Generator, a Discriminator, and a Classifier. 
Mcmt-gan~\cite{ref:Mcmt-gan} proposed an unsupervised modality transferable framework for 3D brain image synthesis. To ensure robust and high-fidelity cross-modal generation, Mcmt-gan introduced three losses (i.e., cycle-consistency loss, bidirectional adversarial loss, and domain adapted loss) and a volumetric manifold regularization. Moreover, a fault-aware discriminator was proposed to make the model coherent with the segmentation task. The cross-modality synthesis tasks include: Proton Density (PD)$\rightarrow$T2, T2$\rightarrow$PD, T1$\rightarrow$T2, and T2$\rightarrow$T1. 
To reflect the texture details of image structure, Ea-GANs~\cite{ref:Ea-GANs} devised an edge-aware generative adversarial network for cross-modality synthesis. Specifically, the edge information was extracted by the Sobel~\cite{ref:Sobel} edge detector and incorporated in the training process of GANs. Two variants were proposed depending on where the edge information was included in the objective function: gEa-GAN (generator) and dEa-GAN (generator and discriminator). 
Both the T1$\rightarrow$FLAIR and T1$\rightarrow$T2 tasks were evaluated. 
Recently, the Transformer~\cite{ref:transformer} was adopted in ResViT~\cite{ref:ResViT} for the first time to conduct multi-modality synthesis. The generator of ResViT has an encoder-decoder structure with multiple Aggregated Residual Transformer (ART) blocks between the encoder and the decoder. The weight-shared ART blocks are able to distill the task-specific information, such as location and context. Three synthesis tasks were conducted: T1,T2$\rightarrow$PD, T1,T2$\rightarrow$FLAIR, and MRI$\rightarrow$CT. 
To tackle the issue of domain-specific deformations in cross-domain synthesis, DiCyc~\cite{ref:R1_p4_brain_crossD} incorporated deformable convolution layer in CycleGAN and designed additional two losses: domain-invariant cycle consistency loss and NMI-based alignment loss. The synthesis tasks were: T2$\rightarrow$PD, PD$\rightarrow$T2, CT$\rightarrow$T2, and T2$\rightarrow$CT.
To tackle the misalignment between the source domain and synthesized images, TPSDicyc~\cite{ref:R1_p5_crossD_synthesis} improved DiCyc with a Spatial Transformation Network (STN) to model the relative deformation. The PD$\rightarrow$T2 and T2$\rightarrow$PD synthesis were evaluated on brain MR images. 

By using cross-modality synthesis, volumetric data for a specific modality can be completed. 
Liao~\emph{et al}.~\cite{ref:Liao_crossmodality} introduced a cross-modality volume completion task to assist in the segmentation of IntraCardiac Echocardiography images. A GAN model named 3D segmentation and completion network (3D-SCNet) was proposed in \cite{ref:Liao_crossmodality} to perform simultaneous completion and segmentation. 
A number of cross-modality completion methods have been proposed to improve Alzheimer's disease diagnosis~\cite{ref:PAN,ref:RevGAN,ref:Lin_crossmodality}. 
PAN~\emph{et al}.~\cite{ref:PAN} utilized the CycleGAN~\cite{ref:un_image_to_image} model to perform 3D PET image imputation from MRI and likely incomplete PET images. Then, the imputed PET and MRI were used by the Landmark-based Multi-Modal Multi-Instance Learning (LM$^3$IL) for Alzheimer's disease classification. 
In 3D-RevGAN~\cite{ref:3D-RevGAN} and Lin~\emph{et al}.~\cite{ref:Lin_crossmodality}, the 3D Reversible GAN~\cite{ref:RevGAN} was adopted to complete the missing PET data for Alzheimer’s Disease diagnosis. Specifically, Lin~\emph{et al}.~\cite{ref:Lin_crossmodality} proposed a two-step method: \textbf{(1)} learn the bidirectional mapping between PET and MRI images with a 3D Reversible GAN, and \textbf{(2)} train a 3D CNN to classify the disease. Reversible GAN has the advantage of only using one generator instead of two in CycleGAN. 

There are some other tasks that are related to the cross-modality/multi-modality synthesis, e.g., quantitative susceptibility mapping (QSM)~\cite{ref:QSMGAN} and Time-of-Flight Magnetic Resonance Angiography (TOF-MRA) generation~\cite{ref:Subramaniam,ref:TOF-MRA}. 
An architecture based on the 3D U-Net was used by QSMGAN~\cite{ref:QSMGAN} to synthesize QSM. 
WGAN with gradient penalty was used to improve the training stability. 
Subramaniam~\emph{et al}.~\cite{ref:Subramaniam} simultaneously generated the TOF-MRA patches along with the corresponding brain vessel segmentation labels. 
To improve efficiency, four variants of the WGAN architecture were considered and the mixed precision technique was applied. 
For the cardiac digital twins task, HDL~\cite{ref:R1_p1_cardiac} proposed a hybrid deep-learning framework consisting of a hybrid UNet for temporal interpolation, a GAN for phase synthesis and a velocity assessment module. For phase synthesis, magnitude images were input to the pixel-to-pixel~\cite{ref:image_to_image} model to synthesize phase images. 
For COVID-19 CT images, a controllable and simultaneous synthesizer (CS$^2$) was proposed in \cite{ref:R1_p2_chest} to generate both images and corresponding segmentation masks for data augmentation. 

\subsubsection{\textbf{Reconstruction/Recovery.}} 
A complex-valued CNN (i.e., $\mathbb{C}$Net~\cite{ref:CNet}) was used to reconstruct under-sampled 3D Late Gadolinium Enhancement (LGE) cardiac MRI data. 
The overall architecture adopted a U-Net structure with several complex-valued components: complex convolutional layer, radial Batch Normalization (BN) layer, down-sampling and up-sampling layers, and complex activation function. $\mathbb{C}$Net achieved better performance compared to real-valued networks and faster speed compared to compressed sensing. 
Zhang \emph{et al}.~\cite{ref:Zhang_recons} used a 2D super-resolution technique to perform 3D Brain MRI super-resolution reconstruction. A super-resolution GAN (i.e., ESRGAN~\cite{ref:ESRGAN}) was applied to 2D brain slices, followed by a Bilinear interpolation to complete the null values. 
ProvoGAN~\cite{ref:ProvoGAN} proposed a progressively volumetric strategy to decompose the 3D volume recovery into sequential 2D cross-sectional mappings. The three rectilinear orientations (axial, coronal, sagittal) were gradually processed in the first, second, and third progression steps. In each step, a 2D GAN was trained based on the previous step results. ProGAN outperformed both volumetric and cross-sectional models on the MRI recovery and synthesis tasks.

\subsection{Segmentation} 


Image segmentation is one of the most important medical imaging applications. 
It can be utilized to either assign corresponding labels to different anatomical structures (e.g., different cardiac chambers) or highlight specific regions of interest (e.g., brain tumor or lesion areas). 
Automatic segmentation liberates medical experts from the time-consuming and tedious task of manually annotating the anatomical structures/disease regions. 
3D volume segmentation offers more detail and anatomical information than 2D image segmentation, which is the focus of this subsection. 
This subsection is organized according to different segmented structures/regions of interest in the brain and heart. 
Hopefully, this will assist researchers in finding their target models more quickly. Table~\ref{tab:segmentation} provides a list of the most relevant methods.

\begin{table}[ht]
\centering
\caption{\label{tab:segmentation}Summary of publications on 3D Segmentation related to the brain or heart.}
\resizebox{1.0\textwidth}{!}{
\begin{tabular}{l|l|l|l|l}\hline
Publication \small{(Year)}    & Organ &Model     & Dataset         & Metrics       \\\hline

Salehi~\emph{et al}.~\cite{ref:Salehi} \small{(2017)} & Brain & U-Net~\cite{ref:UNET} & MSSEG~\cite{ref:MSSEG} & \small{DSC, Sen, Spe, F2, APR} \\\hline 

Myronenko~\cite{ref:Myronenko_segment} \small{(2018)}  & Brain &  VAE~\cite{ref:VAE} & BRATS~\cite{ref:brats18} & Dice, HSD \\\hline 

S3D-UNet~\cite{ref:S3D-UNet} \small{(2018)} & Brain & U-Net~\cite{ref:UNET} & BraTS~\cite{ref:brats18} & \small{Dice, Sen, Spe, HD95} \\\hline 

voxel-GAN~\cite{ref:voxel-GAN} \small{(2018)} & Brain & cGAN~\cite{ref:cond_gans}, U-Net~\cite{ref:UNET} & BRATS~\cite{ref:brats18}, ISLES~\cite{ref:ISLES} & Dice, HSD, Spe, Sen \\\hline 

Mondal~\emph{et al}.~\cite{ref:Mondal} \small{(2018)} & Brain & U-Net~\cite{ref:UNET}, GAN~\cite{ref:gans} & \small{iSeg-2017~\cite{ref:iSeg}, MRBrainS~\cite{ref:MRBrainS}} & DSC, ASD \\\hline 

Yang~\emph{et al}.~\cite{ref:Yang_segment} \small{(2018)} & Heart & GAN~\cite{ref:gans}, U-Net~\cite{ref:UNET} & Own~\cite{ref:Yang_segment} & PSNR, MSE\\\hline 

MuTGAN~\cite{ref:MuTGAN} \small{(2018)} & Heart & GAN~\cite{ref:gans} & Own~\cite{ref:MuTGAN} & ACC, Dice, Sen\\
& & & & Spec, Infarct Size, etc.  \\\hline 

VoxelAtlasGAN~\cite{ref:VoxelAtlasGAN} \small{(2018)} & Heart & cGAN~\cite{ref:cond_gans}  & Own~\cite{ref:VoxelAtlasGAN} & \small{MSD, HSD, Dice, Corr of EF} \\\hline 

Zhang \emph{et al}.~\cite{ref:Zhang_segmentation} \small{(2018)} & Heart & CycleGAN~\cite{ref:un_image_to_image} & Own~\cite{ref:Zhang_segmentation} & Dice, S-score\\\hline 

Liu~\emph{et al.}~\cite{ref:liu_segment} \small{(2019)} & Brain & GAN~\cite{ref:gans}, U-Net~\cite{ref:UNET} & BRATS~\cite{ref:brats18} & Dice \\\hline 

RP-Net~\cite{ref:RP-Net} \small{(2019)} & Brain & U-Net~\cite{ref:UNET} & CANDI~\cite{ref:CANDI}, IBSR~\cite{ref:IBSR} & DSC \\\hline 

DSTGAN~\cite{ref:DSTGAN} \small{(2020)} & Heart & GAN~\cite{ref:gans} & Own~\cite{ref:DSTGAN} & ACC, Dice, Infarct size \\ 
& &  &  & \small{IoU, Per-size, Per-seg} \\\hline 

PSCGAN~\cite{ref:PSCGAN} \small{(2020)} & Heart & GAN~\cite{ref:gans} & Own~\cite{ref:PSCGAN} & \small{SSIM, NRMSE, PSNR, Dice} \\ 
& & & & ACC, Sec, Spe \\\hline 

Yuan~\emph{et al}.~\cite{ref:Yuan_segment} \small{(2020)}& Brain & GAN~\cite{ref:gans} & Decathlon~\cite{ref:Decathlon} & \small{Dice, RAVD, ASSD, MSSD} \\
& &  &  & HD95, Sen, Spe \\\hline  

3D DR-UNet~\cite{ref:3D_DR-UNet} \small{(2020)} & Heart & U-Net~\cite{ref:UNET} & ACDC~\cite{ref:ACDC}, LASC~\cite{ref:LASC} & DSC, HD, ASD\\\hline  

Peng~\emph{et al}.~\cite{ref:peng_segment} \small{(2020)}  & Brain &  U-Net~\cite{ref:UNET} & BraTS~\cite{ref:brats18} & Dice \\\hline 

SASSNet~\cite{ref:SASSNet} \small{(2020)} & Heart & GAN~\cite{ref:gans} & Left Atrium~\cite{ref:Left_Atrium} & \small{Dice, Jaccard, ASD, HD95} \\\hline 

Vox2Vox~\cite{ref:Vox2Vox} \small{(2020)} & Brain & Pixel-to-pixel~\cite{ref:image_to_image} & BRATS~\cite{ref:brats18} & Dice, HD95 \\
& & U-Net~\cite{ref:UNET} & & \\\hline 

Kolarik~\emph{et al}.~\cite{ref:Kolarik} \small{(2021)} & Brain & U-Net~\cite{ref:UNET} & MSSEG~\cite{ref:MSSEG} & Dice, Sen \\\hline 

FM-Pre-ResNet~\cite{ref:FM-Pre-ResNet} \small{(2021)} & Heart & VAE~\cite{ref:VAE}, U-Net~\cite{ref:UNET} & MM-WHS~\cite{ref:MM-WHS} & \small{DSC, Jaccard, SD, HD} \\\hline 

Ullah~\emph{et al}.~\cite{ref:Ullah} \small{(2021)} & Brain & U-Net~\cite{ref:UNET} & BRATS~\cite{ref:brats18} & Dice \\\hline 

MVSGAN~\cite{ref:MVSGAN} \small{(2021)} & Heart & GAN~\cite{ref:gans} & \small{Own~\cite{ref:MVSGAN}, STACOM2011~\cite{ref:STACOM2011}} & Dice, Jaccard, HD \\\hline  

Zhang~\emph{et al}.~\cite{ref:Zhang_segment2} \small{(2021)} & Brain & \small{CycleGAN~\cite{ref:un_image_to_image}, U-Net~\cite{ref:UNET}} & BRATS~\cite{ref:brats18} & Dice, HD95, Sen \\\hline 

DAR-UNet~\cite{ref:DAR-UNet} \small{(2022)} & Brain & GAN~\cite{ref:gans}, U-Net~\cite{ref:UNET} & \small{Vestibular Schwannoma}~\cite{ref:vestibular_vestibular} & Dice, ASD \\\hline 

Bustamante~\emph{et al}.~\cite{ref:Bustamante} \small{(2022)} & Heart & U-Net~\cite{ref:UNET} & Own~\cite{ref:Bustamante} & Dice, HD, ASD, Sen \\
& & & & Precision, miss rate \\\hline 
Xing~\emph{et al}.~\cite{ref:R1_p3_cardiac} \small{(2022)} & Heart & U-Net~\cite{ref:UNET} & Own~\cite{ref:R1_p3_dataset} & MAE, PSNR, SSIM \\ \hline

\multicolumn{5}{l}{Own: The dataset was originally collected and processed by the authors of the paper.}\\
\multicolumn{5}{l}{ACC: accuracy. Sen: sensitivity. Spe: specificity. PPV: positive predictive value. HD95: 95\% Hausdorff distance.}\\
\multicolumn{5}{l}{DSC: dice similarity coefficient. MSD: mean surface distance. HSD: Hausdorff surface distance.}\\
\multicolumn{5}{l}{ASD: average surface distance. EF: ejection fractions. Per-size: percentage of infarct size, Per-seg: percentage of segments.}

\end{tabular}   
}
\end{table}

\subsubsection{\textbf{Brain}}
\paragraph{\textbf{Tumor segmentation.}} 
Brain tumor segmentation has gained great research interest recently~\cite{ref:tumor_survey}, especially for 3D brain volumes~\cite{ref:Myronenko_segment,ref:S3D-UNet,ref:peng_segment,ref:Vox2Vox,ref:Zhang_segment2,ref:voxel-GAN,ref:Yuan_segment,ref:Ullah}. 
Myronenko~\cite{ref:Myronenko_segment} used an encoder-decoder structure for tumor subregion segmentation, which consists of a shared encoder, a segmentation decoder and a reconstruction VAE decoder. S3D-UNet~\cite{ref:S3D-UNet} proposed a separable U-Net architecture that replaces the original 3D convolution with a three-branch separable 3D convolution for brain tumor segmentation. 
Peng~\emph{et al}.~\cite{ref:peng_segment} proposed a multi-scale 3D U-Net architecture to capture the features from different spatial resolutions. The architecture consists of several 3D U-Net blocks, with each block utilizing a 3D separable convolution instead of a standard convolution to reduce the time and memory complexities. 
Vox2Vox~\cite{ref:Vox2Vox} devised a 3D volume-to-volume GAN to segment brain tumors, composed of a U-Net generator to transform images to masks and a PatchGAN~\cite{ref:patchGAN} discriminator for adversarial learning. 
Zhang~\emph{et al}.~\cite{ref:Zhang_segment2} proposed a cross-modality framework based on CycleGAN~\cite{ref:un_image_to_image} to segment brain tumors from multi-modality brain data. Specifically, two sub-modules were designed: a CycleGAN-based Cross-modality Feature Transition module to perform cross-modality synthesis, and a Cross-modality Feature Fusion module to integrate multi-modality features for tumor segmentation. 
The conditional GAN model was used in voxel-GAN~\cite{ref:voxel-GAN} to address the imbalanced tumor segmentation task. 
In order to deal with the imbalance issue, a cGAN that contained a segmentor and discriminator was designed with a weighted adversarial loss. 
Yuan~\emph{et al}.~\cite{ref:Yuan_segment} conducted multi-modal segmentation with only unpaired 3D medical volumes. The segmentation task and auxiliary cross-modality translation tasks were tackled in a unified framework, consisting of modules for modality translation and modality recovery. Each module contained a shared encoder, a segmentation decoder, and a translation decoder. 
Ullah~\emph{et al}.~\cite{ref:Ullah} evaluated a number of MRI image pre-processing techniques for improving and segmenting MRI images. 
Gibbs ringing artifact removal was found to be the best processing technique.

\paragraph{\textbf{Sclerosis lesion segmentation.}} 
To address the data imbalance problem of segmenting sclerosis lesions, 
Salehi~\emph{et al}.~\cite{ref:Salehi} used the U-Net structure and proposed a new generalized loss function called Tversky loss. 
For better imbalanced segmentation, the Tversky loss used both the Dice similarity coefficient (DSC) and F$_1$ score. 
Kolarik~\emph{et al}.~\cite{ref:Kolarik} proposed a transfer learning framework for lesion segmentation in which pre-trained 2D CNN weights were transferred to 3D networks. Specifically, a
Planar 3D U-Net was proposed based on the 3D Planar convolution to conduct lesion segmentation. 
For 2D center axial slice, Basaran~\emph{et al.}~\cite{ref:R3_p5} proposed a foreground-based generative model to synthesize pseudo-healthy images and lesions for downstream segmentation tasks.

\paragraph{\textbf{Brain tissue segmentation.}} 
RP-Net~\cite{ref:RP-Net} segmented the brain tissues into three classes: cerebrospinal fluid (CSF), gray matter (GM), and white matter (WM). RP-Net modified the U-Net structure in \cite{ref:UNET} and introduced both a pyramid pooling module and a recursive residual module. Then, multiple stacks of deep networks were constructed with these modules. 
Mondal~\emph{et al}.~\cite{ref:Mondal} performed the 3D multi-modal brain tissue segmentation task in a few-shot setting where only few labeled data were provided. Specifically, the labeled volumes were input to a U-Net generator to predict the segmentation masks, while the unlabeled volumes underwent a variational autoencoder to perform the reconstruction task.

\paragraph{\textbf{Others.}} 
Four MRI sequences (T1, T1c, T2, FLAIR) were used by Liu~\emph{et al.}~\cite{ref:liu_segment} to segment glioma subregions into three classes: whole tumor (WT), enhancing tumor (ET) and tumor core (TC). 
Specifically, they proposed a GAN-regularized 3D U-Net structure, which uses a pixel-to-pixel scheme to concatenate the input with labels for effective training. 
DAR-UNet~\cite{ref:DAR-UNet} modeled the cross-modality medical image segmentation (e.g., brain structure segmentation) within an Unsupervised Domain Adaptation (UDA) framework. 
The first step was to train multiple GANs to disentangle the style and content of both the source and target images. 
Then, these styles were used to generate diverse target images, which were used to train the segmentation model.

\subsubsection{\textbf{Heart}}
\paragraph{\textbf{Whole heart structure segmentation.}} 
FM-Pre-ResNet~\cite{ref:FM-Pre-ResNet} proposed two novel designs for segmenting the whole heart structure. First, the feature merge residual unit was designed to improve existing pre-activation residual units. Second, 3D VAE was used to perform the volume reconstruction task, which regularized and assisted the segmentation task. 
Zhang \emph{et al}.~\cite{ref:Zhang_segmentation} proposed a cross-modality synthesis approach with unpaired training data to boost the multi-modal volume segmentation performance. 
For the cross-modality synthesis task, two GANs were used to construct the CycleGAN~\cite{ref:un_image_to_image}, and the segmentation task utilized the synthesized data to augment the training data. The model was trained with cycle-consistency loss, adversarial loss, and shape-consistency loss.  
Heart segmentation was accomplished using a multi-stage procedure in 3D DR-UNet~\cite{ref:3D_DR-UNet}: a low-resolution localized map was first generated with a Localization Network, then the high-resolution volume was cropped based on the map and fed into a Segmentation Network to finally generate the segmentation mask. 
Both the Localization Network and the Segmentation Network were built using the 3D Dilated Residual U-Net architecture. 
For the automatic segmentation of the 4D flow MRI volume, Bustamante~\emph{et al}.~\cite{ref:Bustamante} used the U-Net structure instead of the regular 3D volume segmentation. 
With the proposed method, the four cardiac chambers, the aorta, and pulmonary artery were all segmented.

\paragraph{\textbf{Left ventricle (LV) / left atrium (LA) segmentation.}} 
VoxelAtlasGAN~\cite{ref:VoxelAtlasGAN} addressed the 3D LV segmentation task on echocardiography images. 
A new voxel-to-voxel mode of the conditional GANs was devised to include the 3D LV atlas as powerful condition information, addressing issues with limited annotation and low contrast in echocardiography. In addition, a consistent constraint was proposed to enhance the discrimination loss. 
Yang~\emph{et al}.~\cite{ref:Yang_segment} proposed a probability-based method to segment the LV myocardium, avoiding the boundary-moving issues associated with threshold-based methods. 
A 3D U-Net generator was used to propagate the partial 2D slice segmentation to the whole 3D volume, and a 3D CNN discriminator was used to distinguish the ground truth and the generated segmentation masks. 
MVSGAN~\cite{ref:MVSGAN} tackled the short-axis (SAX) view segmentation of 3D cardiac MRI volumes, which was hampered by the sparse spatial structure. Three modules were devised in MVSGAN: \textbf{(1)} a residual adversarial fusion (RAF) module to examine the inter-slice dependency and anatomical priors, \textbf{(2)} a structural perception-aggregation (SPA) module to model the correspondence between segmentation labels and cardiac models, and \textbf{(3)} a joint training objective for multi-task learning. 
SASSNet~\cite{ref:SASSNet} proposed a shape-aware method for semi-supervised 3D LA segmentation~\cite{ref:LA_segment_semi}, incorporating geometric shape constraints into the GANs framework. 
More specifically, the adversarial generation of Signed Distance Map (SDM) serves as an auxiliary task to augment the LA segmentation. The shared segmentation network (generator) adopts a V-Net~\cite{ref:V_net} to simultaneously generate the SDM and segmentation mask. 
Xing~\emph{et al}.~\cite{ref:R1_p3_cardiac} proposed a multi-task attention structure utilizing three UNets to synthesize three-directional CINE multi-slice myocardial velocity mapping (3Dir MVM) and the corresponding left ventricle segmentation.

\paragraph{\textbf{Myocardial infarction (MI) segmentation.}} 
A generative and multi-task learning framework was proposed in MuTGAN~\cite{ref:MuTGAN} for simultaneous quantification and segmentation of myocardial infarction (MI). 
Specifically, the MuTGAN generator consists of two networks: a spatial-temporal feature extraction network (SFEN) to extract the comprehensive spatio-temporal correlations and a joint feature learning network to learn the interaction between segmentation and quantification. The discriminator consists of a task-relatedness network to learn the intrinsic task patterns. 
DSTGAN~\cite{ref:DSTGAN} extended MuTGAN~\cite{ref:MuTGAN} with several enhancements: a new feature learning network (spatio-temporal variation encoder), a conditional GAN, and three iterative task-specific discriminators.

\paragraph{\textbf{Ischemic disease image segmentation.}} 
PSCGAN~\cite{ref:PSCGAN} proposed the first one-stop contrast agent-free framework to simultaneously synthesize and segment the Ischemic Heart Disease (IHD) related cine MR images. The proposed framework is a progressive model integrating three consecutive GAN phases: prior generation, conditional synthesis, and fine segmentation. Each phase uses a similar sequential causal learning framework with two pathways: a spatial perceptual pathway and a temporal perceptual pathway. A synthetic regularization loss and a segmentation auxiliary loss were specially designed to facilitate the model training.

\subsection{Denoising}

\begin{table}[ht]
\centering
\caption{\label{tab:denoising}Summary of publications on 3D Denoising related to the brain or heart}
\resizebox{1.0\textwidth}{!}{
\begin{tabular}{l|l|l|l|l}\hline
Publication \small{(Year)}    & Organ &Model     & Dataset         & Metrics       \\\hline

Wolterink~\emph{et al}.~\cite{ref:Wolterink_denoise} \small{(2017)} & Heart & GAN~\cite{ref:gans} & Harder~\emph{et al}.~\cite{ref:Harder} & PSNR \\\hline

LA-GANs~\cite{ref:LA-GANs} \small{(2018)} & Brain & GAN~\cite{ref:gans} & MCI~\cite{ref:3D_c-GANs} & PSNR, SSIM\\ \hline 

3D c-GANs~\cite{ref:3D_c-GANs} \small{(2018)} & Brain & cGAN~\cite{ref:cond_gans} & MCI~\cite{ref:3D_c-GANs} & \small{PSNR, NMSE, SUV differences} \\\hline

RED-WGAN~\cite{ref:RED-WGAN} \small{(2019)} & Brain & WGAN~\cite{ref:wgan_gp} & IXI~\cite{ref:IXI} & PSNR, RMSE, SSIM, IFC \\\hline

SGSGAN~\cite{ref:SGSGAN} \small{(2022)} & Brain & StyleGAN~\cite{ref:stylegan} & Own~\cite{ref:SGSGAN} & \small{PSNR, SSIM, MAE, U-Net score} \\\hline 

\multicolumn{5}{l}{Own: The dataset was originally collected and processed by the authors of the paper.}\\
\multicolumn{5}{l}{IFC: information fidelity criterion.}\\

\end{tabular}   
}
\end{table}

Artifacts and noise may affect medical images collected in practical clinical settings due to factors such as radiation dose, time constraints, and patient discomfort. 
The purpose of image denoising is to reduce the amount of noise in the original image (e.g., low-dose CT) and recover a high-quality medical image (e.g., full-dose CT). 
Denoising Autoencoders and GANs can naturally be applied for image denoising. 
A list of relevant 3D denoising models is provided in Table~\ref{tab:denoising}.

In an early study, Wolterink~\emph{et al}.~\cite{ref:Wolterink_denoise} proposed the use of GANs in low-dose CT for noise reduction. 
A generator network generated routine-dose CT images from low-dose CT images. A discriminator was used to differentiate between the real and generated routine-dose CT images. The GANs were trained using a voxelwise loss. 
An encoder-decoder and WGAN technique were used to denoise MRI images in RED-WGAN~\cite{ref:RED-WGAN}. 
The residual autoencoder structure was devised for the generator network, while three losses were proposed to train the whole network: adversarial loss, MSE loss, and perceptual loss. 
3D c-GANs~\cite{ref:3D_c-GANs} achieved high-quality (full-dose) PET image synthesis from low-dose CT images using 3D conditional GANs. 
The 3D U-Net was used as a generator, and a concatenated progressive refinement strategy was proposed. 
SGSGAN~\cite{ref:SGSGAN} synthesized full-dose PET images using StyleGAN~\cite{ref:stylegan} and a segmentation U-Net. 
Specifically, a style-based generator converted low-dose PET images to full-dose images, which were then processed by the segmentation network and the discriminator network. 
Based on low-dose PETs and corresponding MRI images, LA-GANs~\cite{ref:LA-GANs} synthesized full-dose PETs. 
First, a local adaptive fusion module was designed to integrate all input modalities into a fused image. Then, a U-Net generator and a discriminator were adversarially trained with the real/fake full-dose PET images.

\subsection{Detection}

\begin{table}[ht]
\centering
\caption{\label{tab:detection}Summary of publications on 3D Detection related to the brain or heart}
\resizebox{1.0\textwidth}{!}{
\begin{tabular}{l|c|c|c|c}\hline
Publication \small{(Year)}    & Organ &Model     & Dataset         & Metrics       \\\hline

Uzunova~\emph{et al}.~\cite{ref:Uzunova_detect} \small{(2019)} & Brain & CVAE~\cite{ref:CVAE} & BRATS~\cite{ref:brats18}, LPBA40~\cite{ref:LPBA40} & Dice, Sen, Spec, AUC \\
& & & & Jaccard \\\hline

MADGAN~\cite{ref:MADGAN} \small{(2021)} & Brain & GAN~\cite{ref:gans} & OASIS-3~\cite{ref:OASIS-3},Own~\cite{ref:MADGAN} & AUC \\\hline

3D MTGA~\cite{ref:3D_MTGA} \small{(2022)} & Heart & U-Net~\cite{ref:UNET}, cGAN~\cite{ref:cond_gans} & Own~\cite{ref:3D_MTGA} & Sen, Spe, Pre, ACC \\
& & & & SSIM, PSNR, MAE, RMSE \\
& & & & DSC, USR, OSR, Jaccard \\\hline

Pinaya~\emph{et al}.~\cite{ref:Pinaya_detect} \small{(2022)} & Brain & VQ-VAE~\cite{ref:VQVAE} & \small{MedNIST~\cite{ref:MedMNIST}, BRATS~\cite{ref:brats18}} & Dice, AUPRC, AUROC  \\
& & \small{Autoregressive Transformer}~\cite{ref:autoregressive_transformer} & WMH~\cite{ref:WMH}, MSLUB~\cite{ref:MSLUB} & FPR80, FPR95, FPR99 \\
& & & UK Biobank~\cite{ref:UKbiobank} &  \\\hline

\multicolumn{5}{l}{Own: the dataset was originally collected and processed by the authors of the paper.}\\
\multicolumn{5}{l}{Sen: sensitivity. Spe: specificity. Prec: precision. ACC: accuracy. FPR: false-positive rates.}\\ 
\multicolumn{5}{l}{DSC: dice similarity coefficient. USR: under-segmentation rate. OSR: over-segmentation rate.}\\ 

\end{tabular}   
}
\end{table}

In medical imaging, detection usually refers to the detection of anomalies or pathologies, such as lesions or tumors. 
A supervised detection method requires large amounts of healthy and anomalous data to effectively differentiate between them. This is often not feasible especially due to the scarcity of anomalous data and data imbalance. 
Therefore, unsupervised detection methods have been proposed in order to use generative models to capture healthy data distribution. A list of relevant 3D detection models is provided in Table~\ref{tab:detection} for the brain or the heart.

MADGAN~\cite{ref:MADGAN} designed a two-step GAN-based medical anomaly detection method to detect different types of brain anomalies at various stages, e.g., Alzheimer’s disease at early and late stages, as well as brain metastasis. The generator uses three consecutive brain MRI slices to generate the next three brain slices, while the discriminator classifies these real/fake next three brain slices. After training, the reconstruction loss (i.e., $L_2$) between the real and reconstructed three slices is computed to detect anomalies. 
A multi-task generative framework was proposed in \cite{ref:3D_MTGA} to detect aortic dissection using non-contrast-enhanced CT (NCE-CT) volumes. The framework consists of a 3D nnU-Net to perform aortas segmentation of NCE-CT volumes and a 3D Multi-Task Generative Architecture (MTGA) to perform CE-CT synthesis, true \& false lumen segmentation, and simultaneous aortic dissection detection. 
Uzunova~\emph{et al}.~\cite{ref:Uzunova_detect} performed unsupervised pathology detection with a 3D CVAE structure. The pathologies were detected by the differences to the learned norm of the healthy image volumes through CVAE reconstruction. 
Pinaya~\emph{et al}.~\cite{ref:Pinaya_detect} applied a Transformer-based model for unsupervised 3D brain anomaly detection and segmentation when a limited amount of data is required. Specifically, a codebook of brain image representations is learned with the VQVAE~\cite{ref:VQVAE} architecture. Then, an Autoregressive Transformer uses the codebook to generate brain volumes in a similar way to PixelRNN~\cite{ref:pixelRNN}. 
Anomalies are considered for Autoregressive Transformer output probabilities lower than a threshold.

\subsection{Registration}

\begin{table}[ht]
\centering
\caption{\label{tab:registration}Summary of publications on 3D Registration related to the brain or heart}
\resizebox{1.0\textwidth}{!}{
\begin{tabular}{l|c|c|c|c}\hline
Publication \small{(Year)} & Organ &Model     & Dataset         & Metrics       \\\hline

VTN~\cite{ref:VTN} \small{(2019)} & Brain & U-Net~\cite{ref:UNET}  & ADNI~\cite{ref:ADNI}, ABIDE~\cite{ref:ABIDE}, ADHD~\cite{ref:ADHD200} & Seg. IoU, Lm. Dist., Time \\\hline

VoxelMorph~\cite{ref:VoxelMorph} \small{(2019)} & Brain & U-Net~\cite{ref:UNET} & \small{OASIS~\cite{ref:OASIS}, ABIDE~\cite{ref:ABIDE}, ADHD~\cite{ref:ADHD200}}  & Dice, Jacobian matrix \\
& & & \small{MCIC~\cite{ref:MCIC}, PPMI~\cite{ref:PPMI}, Harward GSP~\cite{ref:harwardGSP}} &\\
& & & \small{HABS~\cite{ref:HABS}, FreeSurfer Buckner40~\cite{ref:FreeSurfer}} & \\\hline

Deform-GAN~\cite{ref:Deform-GAN} \small{(2020)} & Brain & GAN~\cite{ref:gans} & BRATS~\cite{ref:brats18} & RMSE, Dice \\\hline

Zhu~\emph{et al}.~\cite{ref:Zhu_registration} \small{(2021)} & Brain & U-Net~\cite{ref:UNET} & Mindboggle101~\cite{ref:Mindboggle101}, LPBA40~\cite{ref:LPBA40}, IXI~\cite{ref:IXI} & DSC, HD, ASSD\\\hline

Krebs~\emph{et al}.~\cite{ref:R3_p3} \small{(2021)} & Heart & VAE~\cite{ref:VAE} & Own, ACDC~\cite{ref:ACDC} & Dice, HD \\\hline

Ramon~\emph{et al}.~\cite{ref:Ramon_registration} \small{(2022)} & Brain & GAN~\cite{ref:gans} & ADNI~\cite{ref:ADNI}, NIREP~\cite{ref:NIREP} & DSC \\\hline

TGAN~\cite{ref:TGAN} \small{(2022)} & Brain & GAN~\cite{ref:gans} & BrainWeb~\cite{ref:Brainweb}, Atlas~\cite{ref:ATLAS}, RIRE~\cite{ref:RIRE} & Dice, mean SSIM \\
& & U-Net~\cite{ref:UNET} & & \\\hline

\multicolumn{5}{l}{Own: The dataset was originally collected and processed by the authors of the paper.}\\
\multicolumn{5}{l}{
DSC: dice similarity coefficient. HD: Hausdorff distance. ASSD: average symmetric surface distance.}\\ 

\end{tabular}   
}
\end{table}

The purpose of image registration is to find the spatial correspondence (e.g., affine matrix, deformation field) between an image pair. 
As a result of this correspondence, geometric models like Spatial Transformer Network (STN)~\cite{ref:STN} can be used to warp the image. 
Since manually annotating the corresponding points of image pairs is prohibitively laborious, various unsupervised registration approaches based on generative models (e.g., U-Net~\cite{ref:UNET} or GAN~\cite{ref:gans}) have been proposed.  
Table~\ref{tab:registration} lists some of the most relevant 3D registration methods.

VTN~\cite{ref:VTN} proposed an unsupervised Volume Tweening Network (VTN) for 3D volume registration. VTN was constructed by an end-to-end cascade of registration subnetworks to generate deformation fields between images. The registration subnetwork was implemented with either an affine registration subnetwork or a dense deformable registration subnetwork. An invertibility loss was further incorporated to ensure backward consistency. 
In an end-to-end framework, Zhu~\emph{et~al}.~\cite{ref:Zhu_registration} performed a joint affine alignment and deformable registration. 
The framework contained an affine alignment subnetwork to learn the affine matrix and a deformable registration subnetwork to generate the deformation fields. 
An STN~\cite{ref:STN} was used to warp the volume. 
The VoxelMorph algorithm~\cite{ref:VoxelMorph,ref:VoxelMorph_conf} formulated registration as a mapping between image pairs and deformation fields. A U-Net processes the input pair to get a registration field, which is then used by STN to warp the image. Both the original image volumes and the segmentation masks could be adopted to train the model. 
Deform-GAN~\cite{ref:Deform-GAN} incorporated mono-modal registration and multi-modal registration in an adversarial framework, which consists of a U-Net-like Transformation Network and Generator, as well as a Discriminator. 
The registration model was trained using both a GAN loss and a new local gradient loss. 
Ramon~\emph{et al}.~\cite{ref:Ramon_registration} combined a GAN with the traditional LDDMM~\cite{ref:LDDMM} approach and proposed two model variants for the stationary and non-stationary parameterizations. 
TGAN~\cite{ref:TGAN} proposed a two-stage GAN framework for multi-modal brain registration. The GAN of the first stage extracted the structural representation of the registered image, while the GAN of the second stage estimated the image intensities after registration. 
For spatial-temporal registration of cardiac cine-MRI sequences, Krebs~\emph{et al}.~\cite{ref:R3_p3} proposed a conditional VAE model based on multivariate Gaussian prior and captured the intrinsic motion into a low-dimensional space (i.e., motion matrix). Motion matrix can be used for temporal interpolation, motion transport, motion simulation, and diffeomorphic tracking.

\section{Discussion and Future Directions}
We have surveyed seven medical applications pertaining to 3D generative models for the brain and heart (from Jan 2017 to Oct 2022): unconditional synthesis, classification, conditional synthesis, segmentation, denoising, detection, and registration. 
Among them, segmentation and conditional synthesis have the largest number of publications: 25 and 19, showing the great research interest and importance of these two applications. 
U-Net~\cite{ref:UNET} is the most popular model for segmentation, either used individually or as a generator/encoder network. Image-to-image translation~\cite{ref:image_to_image,ref:un_image_to_image} is naturally compatible with conditional synthesis, and is based on models such as cGAN~\cite{ref:cond_gans}, CycleGAN~\cite{ref:un_image_to_image}, StarGAN~\cite{ref:StarGAN}, etc. 
Unconditional synthesis covers 10 papers, favoring WGAN-GP~\cite{ref:wgan_gp}, VAE~\cite{ref:VAE}, and StyleGAN2~\cite{ref:stylegan2} models. 
In the literature, there are only 6 papers on registration, 5 on denoising, and 4 on detection. 
They are mainly based on the vanilla GAN~\cite{ref:gans} and U-Net~\cite{ref:UNET} models. 
The architecture and model design should consider the application requirements, e.g., CycleGAN~\cite{ref:un_image_to_image} was widely-used for unsupervised learning~\cite{ref:PAN,ref:Zhang_segmentation,ref:Zhang_segment2}, and VAE~\cite{ref:VAE} was adopted for interpreting the classification results~\cite{ref:cardiac_classification,ref:CRT_response}. 
As well as models, we also find that metrics and datasets are usually specific to organs and applications, e.g., ADNI~\cite{ref:ADNI} dataset for brain applications, Dice metric for segmentation-related tasks. 

Although many publications show the effectiveness and application prospects of 3D generative models for medical volumes, this field is still rapidly growing and embraces various potential future directions and techniques. 
The following are five possible future directions.

\subsection{Few-shot Learning} 
Few-shot learning~\cite{ref:few_survey} aims to leverage few available examples to effectively train a deep model. 
It has been extensively researched in computer vision~\cite{ref:few_cv} and natural language processing~\cite{ref:few_nlp}. 
For example, few-shot image generation~\cite{ref:few_generation}, few-shot unsupervised image-to-image translation~\cite{ref:few_translation} have been investigated for natural images. 
As a result of the collection procedure, patient consent, and annotation cost, medical applications usually lack sufficient training data. 
Thus, few-shot learning is naturally suited to medical applications of 3D generative models. 
Using few training samples, Mondal~\emph{et al}.~\cite{ref:Mondal} achieved comparable segmentation results for 3D multi-modal volumes, showing promising prospects for 3D medical applications utilizing few-shot learning methods.

\subsection{Self-supervised Learning} 
Self-supervised learning (SSL)~\cite{ref:SSL} explores the structural and contextual information lying in the data to provide supervision signals for model training. 
Self-supervised learning has the advantage of not requiring manual annotations, saving time and money. 
According to a recent study~\cite{ref:maskAE}, self-supervised learning can be as effective as or even superior to supervised learning in various computer vision tasks. 
For 3D medical applications, the utilization of self-supervised learning will mitigate the burden of tedious and laborious annotation, while still achieving satisfactory performance. 
Joyce \emph{et al}.~\cite{ref:Joyce} trained a 3D cardiac MRI synthesis model with only unlabelled data through self-supervision (reconstruction error). 
Cheng~\emph{et al}.~\cite{ref:SSL_medical} showed that 2D few-shot medical segmentation could be realized with self-supervised learning techniques. 
In the future, we expect more studies to leverage the intrinsic 3D structure of medical volumes as a powerful tool for self-supervised learning.

\subsection{Efficient 3D Network Architecture} 
3D network architecture is another bottleneck for effective 3D generative medical applications, in addition to data scarcity. 
The majority of 3D models for medical volume generation use 3D convolution directly, which is slow and parameter-inefficient. Due to this, existing methods have a small model capacity (e.g., 64 convolution channels~\cite{ref:3d_stylegan} and a small synthesized resolution (e.g., $64\times 64\times 64$~\cite{ref:3D_a_WGANGP})). 
The development of efficient network architectures for 3D generative models is an important and practical research area. 
By using Split\&Shuffle modules in GANs, \cite{ref:split_shuffle} reduced over half of the model parameters, and yet achieved better performance on both heart and brain datasets.
A comprehensive analysis of resource-efficient 3D CNNs was conducted in \cite{ref:resource}. 
Additionally, neural architecture search (NAS)~\cite{ref:NAS} is a promising direction for designing efficient 3D architectures.

\subsection{Denoising Diffusion Probabilistic Model (DDPM)} 
Recent research has shown that DDPMs outperform GANs at generating natural images~\cite{ref:beat_gans,ref:beat_gans2}. 
DDPM models a Markov chain process to convert the simple Gaussian noise into a real data distribution. 
When compared to GANs, it can better cover the overall data distribution and have a stationary training objective. 
As a result, medical applications can also benefit from the DDPM for effective image generation. 
DDPM~\cite{ref:DDPM} was directly used in \cite{ref:diffusion_detect_segment} for unsupervised brain anomaly detection on 2D CT and MRI images. 
In \cite{ref:diffusion_generation}, DDPM was used to estimate the latent code for 4D cardiac MR image generation. 
We believe that future 3D medical volume models can use DDPM as a good alternative to GANs and VAEs.

\subsection{Trustworthy, Explainability, and Potential Bias} 
The pursuit of trustworthy and explainable Artificial Intelligence (AI) is a growing area of research, especially in the field of medical image generation, where diagnostic errors can have severe consequences for patients~\cite{ref:trustworthy}. Recent research has explored creating reliable and explainable AI solutions for both non-imaging medical data synthesis~\cite{ref:R1_p6_nonimaging_trust} and medical image analysis~\cite{ref:explainable}. Given that medical datasets often come from various hospitals, each using different scanning equipment and protocols, models trained on a single dataset can exhibit inherent biases and may not be universally applicable. Recent initiatives~\cite{ref:R1_p6_nonimaging_trust,ref:R1_p7_explainablity} have tackled this issue by harmonizing and fusing data across multiple modalities and centers to minimize bias and enhance explainability. Additional biases may arise from cultural and geographical factors in data collection and model architecture choices. In this survey, we focus primarily on brain and heart applications when exploring possible 3D volume generation models. We anticipate future research will broaden the scope to include more diverse and explainable medical image-generation techniques. 

\section{Conclusion}
This paper reviews three types of generative models and their applications to synthetic 3D medical volume generation. 
A new taxonomy of unconditional and conditional generative models is proposed to cover seven medical applications: unconditional synthesis, classification, conditional synthesis, segmentation, denoising, detection, and registration. 
We have provided a corresponding table with the organ, model, dataset, and evaluation metrics for each application to make it easier for researchers to locate their model of interest. 
Finally, with the hope of further progressing future research in this field, four prospective directions are discussed.

\bibliographystyle{ACM-Reference-Format}
\bibliography{main}

\end{document}